\documentclass[twocolumn,resetfootnote]{aastex701}

\usepackage{pifont}
\shorttitle{LFBOT Late-time Flare Limits}
\shortauthors{Jayaraman et al.}

\defcitealias{ho_tsd_flares}{H23}

\submitjournal{ApJ}
\begin{document}

\title{Constraints on Late-Time Flaring from Luminous Fast Blue Optical Transients using the Transiting Exoplanet Survey Satellite and the Zwicky Transient Facility}

\correspondingauthor{Rahul Jayaraman}

\author[orcid=0000-0002-7778-3117]{Rahul Jayaraman}
\affiliation{Department of Astronomy, Cornell University, Ithaca, NY 14850}
\email[show]{rj438@cornell.edu}  

\author[orcid=0000-0002-9017-3567]{Anna Y. Q. Ho}
\affiliation{Department of Astronomy, Cornell University, Ithaca, NY 14850}
\email{ayh24@cornell.edu}

\author[orcid=0000-0002-9113-7162]{Michael M. Fausnaugh}
\affiliation{Department of Physics and Astronomy, Texas Tech University, Lubbock TX, 79409-1051, USA}
\email{michael.fausnaugh@ttu.edu}

\author[orcid=0000-0002-6786-8774]{Eran Ofek}
\affiliation{Department of Particle Physics and Astrophysics, Weizmann Institute of Science, 76100 Rehovot, Israel}
\email{ofek.eran@gmail.com}

\author[orcid=0000-0001-8472-1996]{Daniel A. Perley}
\affiliation{Astrophysics Research Institute, Liverpool John Moores University, 146 Brownlow Hill, Liverpool L3 5RF, UK}
\email{d.a.perley@ljmu.ac.uk}

\author[orcid=0000-0003-1892-2356]{Ruslan Konno}
\affiliation{Department of Particle Physics and Astrophysics, Weizmann Institute of Science, 76100 Rehovot, Israel}
\email{ruslan.konno@weizmann.ac.il}

\author[orcid=0000-0002-2607-138X]{Martti Kristiansen}
\affiliation{Brorfelde Observatory, Observator Gyldenkernes Vej 7, DK-4340 Tølløse, Denmark}
\email{martti@outinto.space}

% alphabetical below

\author[0000-0001-9152-6224]{Tracy X. Chen}
\affiliation{IPAC, California Institute of Technology, 1200 E. California
             Blvd, Pasadena, CA 91125, USA}
\email{xchen@ipac.caltech.edu}

\author[0000-0002-8262-2924]{Michael W. Coughlin}
\affiliation{School of Physics and Astronomy, University of Minnesota, Minneapolis, MN 55455, USA}
\email{cough052@umn.edu}

\author[0000-0001-5668-3507]{Steven L. Groom}
\affiliation{IPAC, California Institute of Technology, 1200 E. California
             Blvd, Pasadena, CA 91125, USA}
\email{sgroom@ipac.caltech.edu}

\author[0000-0003-3367-3415]{George Helou}
\affiliation{IPAC, California Institute of Technology, 1200 E. California Blvd,
Pasadena, CA 91125, USA}
\email{gxh@ipac.caltech.edu}

\author[orcid=0000-0002-0129-806X]{K.-Ryan Hinds}
\affiliation{Division of Physics, Mathematics, and Astronomy, California Institute of Technology, Pasadena, CA 91125, USA}
\email{khinds@caltech.edu}

\author[0000-0002-5619-4938]{Mansi M. Kasliwal}
\affiliation{Division of Physics, Mathematics, and Astronomy, California Institute of Technology, Pasadena, CA 91125, USA}
\email{mansi@astro.caltech.edu}

\author[orcid=0009-0006-0726-1328]{Zo\"e McGrath}
\affiliation{Astrophysics Research Institute, Liverpool John Moores University, 146 Brownlow Hill, Liverpool L3 5RF, UK}
\email{z.mcgrath@2024.ljmu.ac.uk}

\author[0000-0003-1227-3738]{Josiah N. Purdum}
\affiliation{Caltech Optical Observatories, California Institute of Technology, Pasadena, CA 91125, USA}
\email{jpurdum@caltech.edu}

\author[0000-0001-7357-0889]{Argyro Sasli}
\affiliation{School of Physics and Astronomy, University of Minnesota, Minneapolis, MN 55455, USA}
\affiliation{NSF Institute on Accelerated AI Algorithms for Data-Driven Discovery (A3D3)}
\email{asasli@umn.edu}

\author[0000-0003-1546-6615]{Jesper Sollerman}
\affiliation{Oskar Klein Centre, Department of Astronomy, Stockholm University, AlbaNova, SE-106 91 Stockholm, Sweden}
\email{jesper@astro.su.se}

% \author{friends}
% \affiliation{Additional affiliations here!}
% \email{}

%% Use the \collaboration command to identify collaborations. This command
%% takes an optional argument that is either a number or the word "all"
%% which tells the compiler how many of the authors above the command to
%% show. For example "\collaboration[all]{(DELVE Collaboration)}" wil include
%% all the authors above this command.
%%
%% Mark off the abstract in the ``abstract'' environment. 
\begin{abstract}
The Luminous Fast Blue Optical Transient (LFBOT) AT2022tsd exhibited minutes-timescale optical 
flares in the tens of days following the initial transient event, likely due to
a central engine---either an accreting black hole or a magnetar. In this paper, we use
data from the Transiting Exoplanet Survey Satellite (TESS) and the Zwicky Transient Facility 
(ZTF) to constrain the occurrence of similar flares in the 12 (of 14) known LFBOTs that 
had observational coverage with TESS from tens of days to thousands of days 
after the transient's initial emission. 
We find seven flare-like signals at the locations of four unique LFBOTs; all seven
can likely be attributed to a solar system object (SSO) 
moving through the TESS aperture.
Assuming all seven flares arise from SSOs, for the LFBOT 
AT2024qfm we rule out flaring with a similar timescale
(40--65\,d) and luminosity ($\nu L_\nu\sim10^{43}$\,erg\,s$^{-1}$) as in AT2022tsd, while for
AT2022tsd itself we rule out flares between 380--430\,d after the initial
transient that were as luminous as the earlier flares. This observation suggests
that the engine power in AT2022tsd declined or shut off on a timescale of
hundreds of days. We also find that there is no late-time activity detectable in TESS thousands
of days after the prototype LFBOT, AT2018cow. 
We discuss our constraints on the duty cycle of such flaring and then
present estimates for the number of minutes-duration flares 
detectable with ongoing and upcoming high-cadence ($\ll1$\,d)
wide-field surveys.
\end{abstract}

%% Keywords should appear after the \end{abstract} command. 
%% The AAS Journals now uses Unified Astronomy Thesaurus (UAT) concepts:
%% https://astrothesaurus.org
%% You will be asked to selected these concepts during the submission process
%% but this old "keyword" functionality is maintained in case authors want
%% to include these concepts in their preprints.
%%
%% You can use the \uat command to link your UAT concepts back its source.

% \keywords{\uat{Galaxies}{573} --- \uat{Cosmology}{343} --- \uat{High Energy astrophysics}{739} --- \uat{Interstellar medium}{847} --- \uat{Stellar astronomy}{1583} --- \uat{Solar physics}{1476}}

\section{Introduction} 

All-sky optical surveys have led to the discovery of new classes of transients
that rise and decay on timescales of a few days, compared to the tens 
of days associated with typical core-collapse supernovae 
(e.g., \citealt{inserra_lfbot_review}). Of particular interest
is the empirical class of Luminous Fast Blue Optical Transients (LFBOTs), 
peaking at absolute magnitudes $-22\leq M \leq -20$ (e.g., 
\citealt{ho_koala,perley_xnd,gutierrez_css161010}).
Unlike typical core-collapse supernovae, LFBOTs exhibit luminous 
emission across the electromagnetic spectrum (e.g., \citealt{margutti_cow_2019}). These
transients bear
marked similarities to engine-powered events such as gamma-ray bursts 
(e.g., \citealt{coppejans_css161010}) and 
tidal disruption events \citep{tsuna_lu_stellar_tde}, as well as to 
interaction-powered supernovae (e.g., of Type Ibn; \citealt{lfbot_ibn_interaction}).

The first LFBOT to be discovered and identified as such, AT2018cow 
\citep{prentice_cow_2018}, had an optical light curve that peaked within a 
few days and decayed to half of maximum light after $\sim$\,3\,d. AT2018cow also had
a persistently blue color and fast ejecta ($v\sim0.1c$; 
\citealt{perley_cow_2019,ho_cow_2019, margutti_cow_2019, nayana_chandra_cow_2021}). An 
estimate of the LFBOT\footnote{In this paper, ``LFBOT'' refers to
transients with optical light curves similar to AT2018cow \citep{sevilla_lfbot_sample}.} rate using the Zwicky Transient Facility (ZTF), 
a ground-based optical transient survey \citep{bellm_ztf, graham_ztf,dekany_ztf_os} suggests that 
these events are rare, being $\sim$\,0.001--0.01\% as common as nearby core-collapse supernovae
\citep{perley_wpp}. As a result, only 14 have been discovered (see Table \ref{tab:fbot_info}).

Numerous other transients have been found to be fast and blue, but their connections 
to LFBOTs are unclear because they were found in archival surveys and therefore have 
limited follow-up observations (e.g. \citealt{drout_ps1_fbot,tanaka_subaru_blue,
arcavi_snls_fbot,pursiainen_fbot_des}). Alternatively, they have been spectroscopically 
classified as known types of core-collapse supernovae, with light curves 
powered by shock interaction (e.g., \citealt{ofek_2010_ptf09uj,ho_fbot_ztf_2023}). 
In contrast, the observational characteristics of LFBOTs, including their long-lived luminous 
X-ray and radio emission, suggest that they are powered at least in part 
by a central engine \citep{ho_cow_2019,margutti_cow_2019,nayana_chandra_cow_2021,yao_mrf,bright_xnd,matthews_tsd,migliori_cow}. Further evidence for a central engine is the detection of
a tentative (3.7-$\sigma$) quasi-periodic oscillation in X-ray emission from
AT2018cow \citep{pasham_cow_central_object} and a luminous late-time UV source 
at the location of AT2018cow \citep{sun_hst_uv_cow,chen_hst_uv_cow_ii,inkenhaag_hst_uv_cow}. However, 
recent modeling of AT2018cow \citep{govreen_segal_model} has proposed that
a shock propagating through an aspherical circumstellar medium could also 
explain much of the observed early ($t < 40$\,d) multi-wavelength emission. 

While the mechanism of AT2018cow's emission remains debated, the unambiguous
detection of late-time, persistent, rapid flaring from the LFBOT AT2022tsd \citep[hereafter \citetalias{ho_tsd_flares}]{ho_tsd_flares},
provided strong support for a central engine in at least some LFBOTs.
The minutes-duration flares emitted from AT2022tsd tens of days
after the initial transient radiated luminosities up to 
$\nu L_\nu \sim 10^{44}$\,erg\,s$^{-1}$ (spherical equivalent).
These observations enabled constraints on the radius of 
the flare-emitting region, which was found to be much smaller than the initial 
transient's photospheric radius. To date, AT2022tsd
is the only LFBOT for which such rapid, luminous, 
long-lasting flares have been detected. It
remains unknown if this flaring behavior is universal among LFBOTs, or whether it occurs in other transient types.
Searches for flaring in the LFBOT AT2024wpp \citep{lebaron_wpp,nayana_wpp,pursiainen_wpp_spherical, perley_wpp}
using the Large Array Survey Telescope
(LAST; \citealt{ofek_last}) yielded non-detections \citep{ofek_wpp_limits}.
One hypothesis for these non-detections is that the flare
emission is beamed, compared to the thermal, more isotropic
emission from the initial transient.

% While there is now evidence for the presence of a central object 
% in at least two LFBOTs (AT2018cow and AT2022tsd), 
% the geometry of the ejecta from the initial explosion remains poorly constrained.
% Observations of AT2024wpp suggest that LFBOTs' ejecta is
% aspherical \citep{nayana_wpp,perley_wpp,lebaron_wpp}. 
% Such a result would imply a shock propagating through an aspherical, possibly disk-like, ejecta 
% geometry (agreeing with the modeling in \citealt{margutti_cow_2019} 
% and \citealt{govreen_segal_model}). 

Targeted searches for LFBOT flares
require significant amounts of telescope time to 
achieve the necessary temporal resolution and coverage (the duty cycle is low), 
and are not a guaranteed success. Moreover, if a detection is made,
the integration times of existing detectors may not be able to capture very
short-timescale variability; variability between 30\,s exposures was observed in 
AT2022tsd. The best (and often only) method to search for flares is using
minutes-cadence wide-field surveys---such as the Large Array
Survey Telescope \citep{ofek_last,ofek_last_image_processing,last_benami}, the Argus Array \citep{argus_array}, and the Transiting Exoplanet Survey Satellite (TESS; \citealt{ricker_tess}). 
Days-cadence surveys such as ZTF,
which has found most known LFBOTs, could
identify single-epoch flux excesses in LFBOTs that imply a flaring state
\citepalias{ho_tsd_flares}.

% In most cases, searches in wide-field survey data
% represent the only available option to search for these flares. 
Moreover, it is crucial
to search for these flares using detectors sensitive
to redder emission---\citetalias{ho_tsd_flares} found that the
flares from AT2022tsd had a $u$--$I$ color of 1.4, 
as opposed to its initial blue color.
This observation, combined with a high brightness temperature 
($\gtrsim$$10^{10}$\,K),
suggests that these flares are nonthermal.
TESS's bandpass has a pivot wavelength similar to that of 
Cousins $I$ but is twice as wide \citep{ricker_tess},
straddling the optical and near-infrared (600--1\,000\,nm), making it an excellent
tool to search for these flares.

In this paper, we use existing TESS and ZTF data 
to constrain the existence, timescale, duty cycle,
and overall rates of late-time flaring in all the currently-known 
LFBOTs. TESS's survey strategy---repeating fields on years-long 
timescales---also allows us to
constrain the recurrence timescale, if any, of these flares.
Our work sets limits on the occurrence of flares
in 12 of 14 currently-known LFBOTs. Section \ref{sec:observations} discusses the TESS
data (and our difference imaging approach) and the ZTF data used as part of our analysis.
Section \ref{sec:results}
presents our limits on the brightness of putative flares, and
Section \ref{sec:discussion} estimates these flares' duty cycle and 
evaluates our ability to discover them using ongoing
and upcoming high-cadence (minutes--hours) wide-field surveys in ground and space.

\begin{table*}
\caption{Information about all 14 LFBOTs detected to date and 
observational details with TESS. Redshift values, coordinates, and discovery epochs
are taken from published 
values in \citet{sevilla_lfbot_sample} unless indicated otherwise, and discovery
epochs are taken from the Transient Name Server (TNS; \citealt{tns}). 
Barycentric corrections were made to the time of discovery (from ZTF, at Palomar Observatory), and
then rewritten in the TESS Barycentric Julian Date format: BJD--2\,457\,000.}
\centering
\begin{tabular}{lcrrclc}
\hline
\hline
\multicolumn{1}{c}{Identifier} &
\multicolumn{1}{c}{Nickname} &
\multicolumn{1}{c}{Right Ascension} &
\multicolumn{1}{c}{Declination} &
\multicolumn{1}{c}{Discovery Epoch} &
\multicolumn{1}{c}{$z$} &
\multicolumn{1}{c}{TESS Sectors} \\
\multicolumn{2}{c}{} &
\multicolumn{1}{c}{(hh mm ss)} &
\multicolumn{1}{c}{(dd mm ss)} & 
\multicolumn{1}{c}{(BJD--2457000)} &
\multicolumn{1}{c}{} \\
\hline
CSS161010$^*$ & & 04h 58m 34.00s & $-08^\circ$ 18m 03.00s & 671.98366 & 0.033& [{\bf 5}, {\bf 32}, {\bf 98}, {\it 109--110}] \\
AT2018cow$^*$ & ``Cow'' &16h 16m 00.22s & +22$^\circ$ 16m 04.91s & 1285.94556 & 0.014 & [{\bf 25}, {\bf 51}, {\bf 52}, {\bf 78}, {\it 117}] \\
AT2018lug$^*$ & ``Koala'' & 02h 00m 15.19s & +16$^\circ$ 47m 57.30s & 1373.91246& 0.271& [{\bf 42--43}, {\bf 70}, {\bf 71}] \\
AT2020mrf$^*$ & & 15h 47m 54.16s & +44$^\circ$ 39m 07.41s & 2015.87459& 0.135 & [23--24, {\bf 50--51}, {\bf 77--78}] \\
AT2020xnd$^*$ & ``Camel'' & 22h 20m 02.03s & $-02^\circ$ 50m 25.30s & 2134.67579& 0.243& [{\bf 42}, {\bf 55}, {\bf 92}] \\
AT2022tsd$^*$ & ``Tasmanian & 03h 20m 10.86s & +08$^\circ$ 44m 55.63s & 2829.97633& 0.256& [4, 31, 42--44, {\bf 70--71}] \\
& Devil''& & & \\
AT2022abfc$^*$ & & 04h 51m 19.20s & $-26^\circ$ 58m 41.60s & 2904.84850& 0.212& [32, {\bf 98}, {\it 106--109}] \\
AT2023fhn & ``Finch'' & 10h 08m 03.82s & +21$^\circ$ 04m 26.95s & 3044.70826& 0.240&  [21, 45--46, 48, {\bf 72}] \\
AT2023hkw & ``Hawk'' & 10h 42m 17.75s & +52$^\circ$ 29m 19.33s & 3065.70049& 0.343& [21, 48, {\bf 75}] \\
AT2023vth & & 17h 56m 34.40s & +08$^\circ$ 02m 37.32s & 3235.61473& 0.075& [{\bf 80}, {\it 118}] \\
AT2024qfm & & 23h 21m 23.46s & +11$^\circ$ 56m 32.04s & 3518.85066& 0.227& [56, {\bf 83}] \\
AT2024aehp & & 08h 21m 07.47s & +28$^\circ$ 44m 22.30s & 3663.86281& 0.172& [44--47, 71--72] \\
AT2024wpp & ``Whippet'' & 02h 42m 05.50s & $-16^\circ$ 57m 22.90s & 3579.87606& 0.087& [4, 31, {\bf 97}, {\it 108}] \\
AT2026dbl & ``Dibbler'' & 09h 01m 17.37s & +18$^\circ$ 36m 07.73s & 4085.85565 & 0.190$^a$ & [44--46, 72]\\
\hline
\end{tabular}
\tablecomments{Sources with $*$ indicate that limits on flares (or detections in the case of AT2022tsd) 
were presented in \citetalias{ho_tsd_flares}, using data from ZTF and other surveys. 
TESS sectors (in the last column) written in \textbf{bold} indicate that those periods 
of observation occurred after the Discovery Epoch given in the fourth column. \textit{Italicized} sectors
indicate future observations that are part of TESS's Cycle 9 observations, 
starting in 2026 September. 
(a) \citet{wise_26dbl}; Z. McGrath et al., in prep.}
\label{tab:fbot_info}
\end{table*}

\section{Observations}
\label{sec:observations}

\subsection{TESS Data and Difference Imaging}
\label{subsec:tess_phot}

\begin{figure*}
    \centering
    \includegraphics[width=\linewidth]{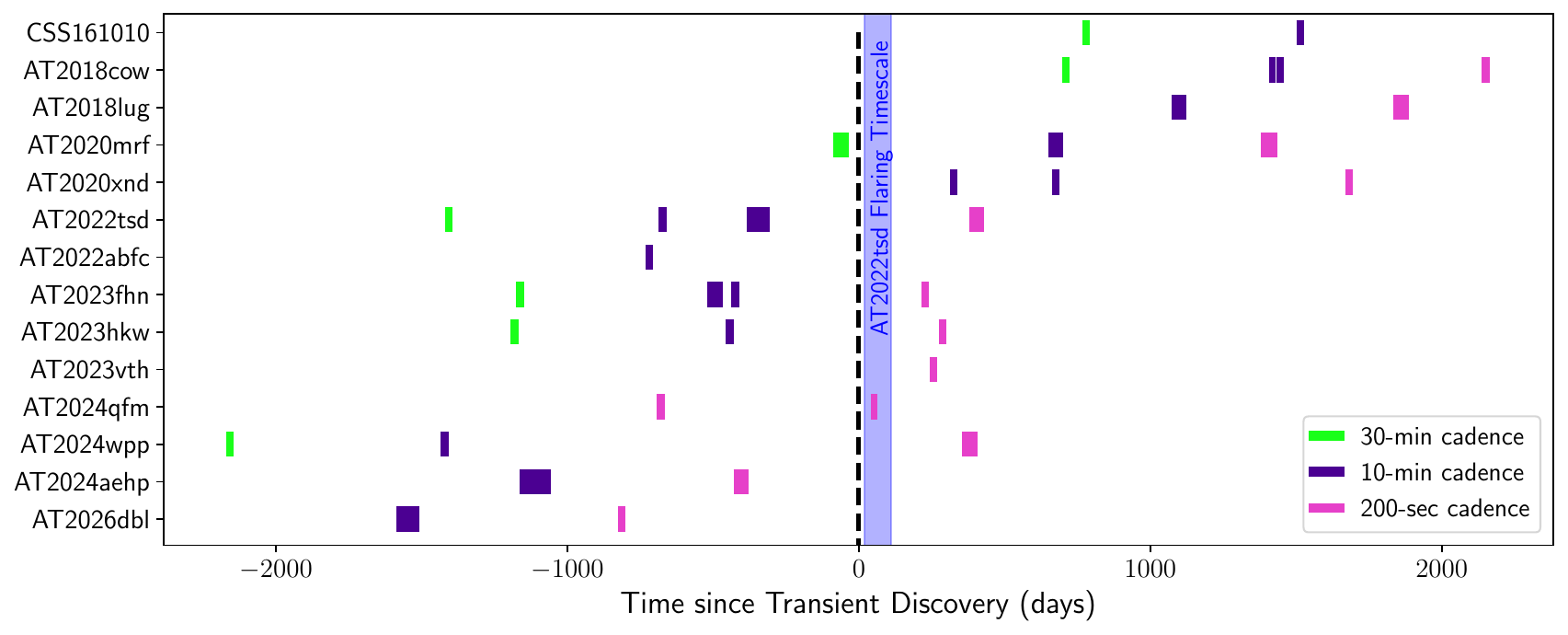}
    \caption{TESS observations of the 14 known LFBOTs compared to their discovery epochs 
    (black dotted line). The different observational cadences (1800s, 600s, 200s)
    are shown in different colors (light green, purple, pink, respectively). 
    We highlight the limits on flaring in AT2024qfm obtained through TESS
    observations that occurred on the same timescale as the flaring in AT2022tsd.
    Two of the 14 LFBOTs (CSS161010 and AT2018cow) occurred prior to the start of TESS observations.}
    \label{fig:obs_timeline}
\end{figure*}

The primary goal of TESS is to find transiting exoplanets around
nearby stars \citep{ricker_tess}. However, its high cadence (hundreds
of seconds) and large field of view (2304\,deg$^2$) make it amenable 
to studying short-timescale transient behavior. At the start of the 
TESS Mission (2018 July--2020 July), the observational cadence of 
TESS's full-frame images (FFIs) was 1800\,s. During Years 3--4 of the
mission (2020 July--2022 September), the cadence was reduced to 600\,s,
and since Year 5 of the mission (2022 September--present), the FFI 
cadence has been 200\,s. TESS's observing strategy for the first several
years of its mission was to cover over 85\% of the sky every two years by
staring at a given field of view (FOV) for roughly a month at a time.
Figure \ref{fig:obs_timeline} shows the times during which TESS observed
each LFBOT, relative to its initial detection.

We extracted TESS light curves at the location of the known LFBOTs (given in Table
\ref{tab:fbot_info}) for every 
observational Sector in which TESS observed their positions. To do so, we performed forced
photometry on difference images created from the calibrated FFIs output by the 
TESS Image CAlibrator (TICA) pipeline \citep{fausnaugh_tica}. These difference 
images are constructed using ISIS \citep{alard_lupton_isis}. In short, each image was
subtracted with respect to a reference image constructed from 20 low-background 
FFIs; systematic errors were fixed through the use of a spatially variable smoothing
kernel that aimed to match the reference image to an individual FFI. Further details
about the full difference imaging procedure can be found
in \citet{faus_tess_ia_2021,faus_tess_ia_2023}. We then performed forced
PSF-fitting photometry on these difference images. Light curves were flux-calibrated 
by first calculating the Vega zero-point in the TESS bandpass
using the CALSPEC Vega model \citep{bohlin_vega}. We found $T = 0$ to correspond to
a flux ($F_\nu$) of 2583\,Jy. We then calculated the TESS magnitude $T$ using
\begin{equation}
    T = -2.5\log_{10}\left(\frac{N}{t\times 0.99 \times 0.8}\right) + 20.44,
\end{equation}
and then converted this to Jy using the expression $F_\nu = 2583$\,Jy\,$\times10^{-T/2.5}$.
Here, $N$ represents the total observed counts from a source, 0.99 estimates the frame
transfer efficiency, 0.8 is the cosmic-ray mitigation factor, and 20.44 is the
TESS zero-point corresponding to 1 e$^-$\,s$^{-1}$.
More details about this flux calibration technique,
as well as TESS's onboard cosmic-ray mitigation algorithm and its effects on
short-timescale variability, can be found in \citet{jayaraman_grb_2024},
as well as the TESS Instrument Handbook\footnote{\url{https://archive.stsci.edu/missions/tess/doc/TESS_Instrument_Handbook_v0.1.pdf}}.
We convert TESS Vega magnitudes to AB magnitudes using a
constant offset (calculated in \citealt{jayaraman_grb_2024}): $T_{\rm AB} = T_{\rm Vega} + 0.3697$. 
The TESS light curves for all observations of LFBOT locations are given in Table \ref{tab:tess_lc}.

\begin{table*}
\caption{TESS light curves of the LFBOTs analyzed in this work. The
light curves are given for the full sector, for all sectors in which the LFBOT
fell. We report the limiting magnitudes in Table \ref{tab:constraints} for flares and
Table \ref{tab:precursor_limits} for precursors. In this table,
we include the time, differential flux, and background estimates from our
photometric pipeline described in Section \ref{subsec:tess_phot}. The background estimate
consists of three values: (1) Local, measuring the local background; (2) Model,
encapsulating the potential systematic errors inherent in the background correction; and
(3) Residual, enumerating the remaining background in the photometric aperture after subtraction
of the background (Local + Model) and measured Flux. Further information about the background modeling
can be found at \url{https://tess.mit.edu/public/tesstransients/pages/readme.html}.}
\centering
\begin{tabular}{cccccccc}
\hline
\hline
\multicolumn{1}{c}{\textbf{Sector}} &
\multicolumn{1}{c}{\textbf{Identifier}} &
\multicolumn{1}{c}{\textbf{Time}} &
\multicolumn{2}{c}{\textbf{Photometry}} &
\multicolumn{3}{c}{\textbf{Background}} \\
\multicolumn{2}{c}{} &
\multicolumn{1}{c}{} &
\multicolumn{1}{c}{Differential Flux} &
\multicolumn{1}{c}{Uncertainty} &
\multicolumn{1}{c}{Local} &
\multicolumn{1}{c}{Model} & 
\multicolumn{1}{c}{Residual} \\
\multicolumn{2}{c}{} &
\multicolumn{1}{c}{(BJD--2\,457\,000)} &
\multicolumn{2}{c}{(ct s$^{-1}$)}&
\multicolumn{3}{c}{(ct s$^{-1}$)} \\
\hline 
5 & CSS161010 & 1438.01566 & 0.0527 & 1.0766 & $-11.5339$ & 0.2296 & 1.280187 \\
5 & CSS161010 & 1438.03650 & 3.1743 & 1.0485 & $-12.3155$ & 0.3522 & 0.390901 \\
... & ... & ... & ... & ... & ... & ... & ...  \\
\hline
\end{tabular}
\label{tab:tess_lc}
\tablecomments{The entirety of this table is available in machine-readable
format.}
\end{table*}

\begin{figure*}
    \centering
    \includegraphics[width=\linewidth]{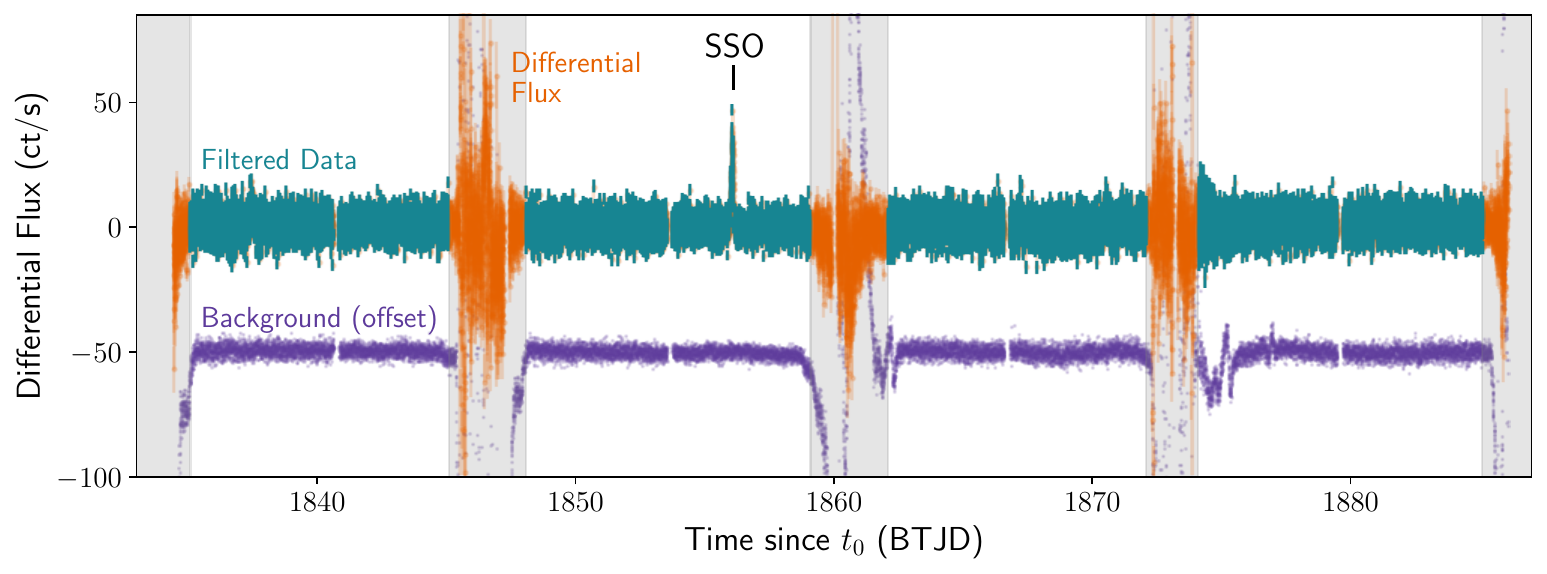}
    \caption{Light curve for AT2018lug from TESS Sectors 70 and 71 showing the 
    differential flux (orange), the filtered data points used in our
    analysis (turquoise), and the backgrounds from scattered light (purple). 
    The background is offset for clarity. 
    The scattered light backgrounds peak toward the end of a given
    orbit (each Sector is two TESS orbits). A solar system object
    passing through the aperture is indicated; for further details, 
    see Section \ref{subsec:ssos} and Table \ref{tab:ssos}. There is no
    evidence of flaring here.}
    \label{fig:tess_lc}
\end{figure*}

An example late-time TESS light curve for an LFBOT is shown in Figure
\ref{fig:tess_lc}. This light curve demonstrates both a large source of
false positives (solar system objects, or SSOs, passing through the 
photometric aperture) and a large source of systematic error---scattered light.
In particular, scattered light from the Earth and the Moon (when they
are above the sunshade) leads to elevated, time-variable
backgrounds, especially toward the 
end of an orbit. To mitigate the effect of scattered light on our 
analysis, the light curves were visually inspected to mask out 
periods of time with backgrounds arising from significant 
scattered light (see Figure \ref{fig:tess_lc}).
We note that background values are centered at 0 in these light curves
because they represent the local backgrounds in the 
difference image, after subtracting the reference image.

We corrected for extinction using the maps of 
\citet{schlafly_extinction}. $A_V$ values were obtained from the NASA
Extragalactic Database extinction calculator
and were corrected using the prescription presented in \citet{jayaraman_grb_2024},
that $A_{\rm TESS} \approx 0.6470\,A_V$. These values were used to correct the
upper limits on flux and luminosity in Tables \ref{tab:constraints}--\ref{tab:precursor_limits}.

\subsection{Zwicky Transient Facility}

To access photometry from ZTF, we use the {\tt Fritz} API, an
instance of {\tt SkyPortal} \citep{skyportal_1,skyportal_2},
which allows us to download all photometry for a given target 
from ZTF, and perform forced photometry on the ZTF difference
images \citep{ztf_diff_infra,masci_ztf_fp_ii}. 
We performed forced photometry at the site 
of each of the 6 LFBOTs in Table \ref{tab:fbot_info} that were not
already searched for flaring in \citetalias{ho_tsd_flares} to collect
all observations (including non-detections) until 30 
March 2026. We then filtered the data to search for any 
detections hundreds of days after the initial detection of the
transient (this epoch is henceforth referred to as $t_0$). For any detection,
we inspected the corresponding images to determine
whether this was genuine. Further details 
are given in Section \ref{sec:results} and Appendix \ref{app:stat_tests}.

\subsection{Other photometry}
\label{subsec:other_phot}
{\tt Fritz} also enables users to
upload photometric and spectroscopic observations that were 
collected as part of follow-up efforts, as well as perform forced
photometry on difference images from
the Asteroid Terrestrial-impact Last Alert System (ATLAS; 
\citealt{tonry_atlas}). We performed ATLAS forced photometry
for all the LFBOTs to ensure maximal coverage, especially
for epochs that were not observed by ZTF due to
weather or other constraints. This functionality relies upon the online ATLAS
forced photometry API, as described in 
\citet{smith_atlas_science_server} and \citet{shingles_atlas_forced_phot}.

\section{Results}
\label{sec:results}

An initial visual inspection of the available light curves from TESS and ZTF yielded
seven detections of flux excesses that resembled flaring behavior, of which all seven were
confirmed to be solar system objects (SSOs) through inspection of the TESS images
and comparisons with the Sky Body Tracker service \citep{berthier_skybot}. Forced photometry
with ATLAS revealed a single detection (with a 4.6-$\sigma$ significance)
for AT2022abfc roughly 380 days after the first detection of the transient, 
though the raw difference image from ATLAS did not show clear evidence of 
a source. Further properties of this detection are discussed in the Appendix.

In addition, we also searched for any precursor emission at the location of these transients in both
TESS and ZTF, and found no evidence for any pre-LFBOT emission at any of these 
locations that was comparable to the peak luminosity of the transient. 
Our results, including upper limits on the 
luminosity of putative flares, are given in Tables \ref{tab:constraints}--\ref{tab:precursor_limits}.
All magnitudes reported are in the AB system, and all calculations involving redshifts will
assume a {\tt Planck18} cosmology as implemented in {\tt astropy} 
\citep{planck_cosmology}, with $H_0=67.66$\,km\,s$^{-1}$\,Mpc$^{-1}$ and $\Omega_M=0.31$.

\begin{table*}[h]
\caption{Results of our searches for LFBOT flares in TESS that occurred after the transient, with limiting magnitudes and
upper limits on the luminosity ($\nu L_\nu$). For contiguous sectors, we shifted the light curves to have a common baseline of zero and 
then analyzed them jointly. Sectors 1--26 had a cadence of 1800\,s, 27--55 had a cadence of 600\,s, and 
56 onward have a cadence of 200\,s. The reported magnitudes and luminosities are corrected for Galactic extinction, as described in 
Section \ref{subsec:tess_phot}.}
\centering
\begin{tabular}{lccllcccl}
\hline
\hline
\multicolumn{1}{c}{\bf Identifier} &
\multicolumn{1}{c}{{\bf LFBOT Init-}} &
\multicolumn{3}{c}{\bf TESS Observation Details} &
\multicolumn{2}{c}{\bf 200\,s cadence limits} &
\multicolumn{2}{c}{\bf 2\,h binned limits} \\
\multicolumn{1}{c}{} &
\multicolumn{1}{c}{{\bf ial Peak $\nu L_\nu$}} &
\multicolumn{1}{c}{Sector} &
\multicolumn{2}{c}{Times} &
\multicolumn{1}{c}{Magnitude} &
\multicolumn{1}{c}{Luminosity} &
\multicolumn{1}{c}{Magnitude} &
\multicolumn{1}{c}{Luminosity} \\
\multicolumn{1}{c}{} &
\multicolumn{1}{c}{(erg\,s$^{-1}$)} &
\multicolumn{1}{c}{} &
\multicolumn{1}{c}{Start} &
\multicolumn{1}{c}{End} &
\multicolumn{1}{c}{$T_{\rm AB}$} &
\multicolumn{1}{c}{(erg\,s$^{-1}$)} &
\multicolumn{1}{c}{$T_{\rm AB}$} &
\multicolumn{1}{c}{(erg\,s$^{-1}$)}\\ 
\hline
& & 5 & $t_0 + 766.04$ & $t_0 + 791.81$ & {18.89} & $<1.04\times10^{42}$ & {19.37} & $<6.69\times10^{41}$ \\
CSS161010 & $4.97\times10^{43}$ (a) & 32 & $t_0 + 1502.28$ & $t_0 + 1528.25$ & {18.20} & $<1.95\times10^{42}$ & {19.32} & $< 6.97\times10^{41}$ \\
 & & 98 & $t_0 + 3322.02$ & $t_0 + 3344.91$  & {17.45}  & $<3.91\times10^{42}$ & {19.31} & $< 7.04\times10^{41}$ \\
 \hline 
 & & 25 & $t_0+697.72$ & $t_0+723.34$  & {18.45} & $<2.73\times10^{41}$ & {19.05} & $<1.57\times10^{41}$ \\
AT2018cow & $4.29\times10^{43}$ (b)& 51--52 & $t_0 + 1407.01$ & $t_0 + 1457.14$ & {18.25} & $<3.27\times10^{41}$ & {19.42} & $<1.11\times10^{41}$   \\
 & & 78 & $t_0+2148.33$ & $t_0+2163.57$  & {17.57} & $<6.17\times10^{41}$ & {19.79} & $<7.93\times10^{40}$ \\
\hline 
AT2018lug & $4.71\times10^{43}$ (c) & 42--43  & $t_0 + 1073.79$ & $t_0 + 1124.98$ & {18.16} & $<1.84\times10^{44}$ & {19.46} & $< 5.54\times10^{43}$ \\
 & & 70--71 & $t_0 + 1834.45$ & $t_0 + 1886.06$ & {17.65} & $<2.96\times10^{44}$ & {19.55} & $< 5.10\times10^{43}$\\ 
\hline
 AT2020mrf & $2.71\times10^{43}$ (d) & 50--51 & $t_0 + 649.40$ & $t_0 + 701.66$ & {18.30} & $<3.45\times10^{43}$ & {19.54} & $<1.10\times10^{43}$ \\
  & & {77-78} & $t_0 + 1379.61$ & $t_0 + 1433.64$ & {17.89} & $< 5.02\times10^{43}$ & $\sim19.5$ & $\lesssim1\times10^{43}$ (e) \\ 
  \hline
 & & {42} & $t_0 + 313.03$ & $t_0 + 338.49$ & {18.02} & $<1.63\times10^{44}$ & {18.71} & $<8.68\times10^{43}$ \\ 
 AT2020xnd & $4.71\times10^{43}$ (f) & 55 & $t_0 + 662.44$ & $t_0 + 689.60$ & {18.09} & $<1.53\times10^{44}$ & {19.38} & $<4.66\times10^{43}$\\
 & & {92} & $t_0 + 1669.33$ & $t_0 + 1692.63$ & {17.41} & $<2.88\times10^{44}$ & {19.52} & $<4.12\times10^{43}$ \\
 \hline 
  AT2022tsd & $6.81\times10^{43}$ (g)& 70--71 & $t_0 + 379.03$ & $t_0 + 429.53$$^g$ & {17.26} & $<3.68\times10^{44}$ & {19.11} & $<6.70\times10^{43}$ \\ 
  \hline
  AT2022abfc & $5.16\times10^{43}$ (h) & 98 & $t_0+1083.65$ & $t_0 + 1112.04$ & {17.94} & $<1.29\times10^{44}$ &{20.39} & $<1.37\times10^{43}$ \\ 
  \hline
  AT2023fhn & $1.08\times10^{44}$ (h) & 72 & $t_0+215.47$ & $t_0+240.88$ & {17.57} & $<2.41\times10^{44}$ & {19.56} & $<3.87\times10^{43}$ \\ 
  \hline 
  AT2023hkw & $6.80\times10^{43}$ (h) & 75 & $t_0+274.09$ & $t_0+301.79$ & {17.89} & $<4.06\times10^{44}$ & {19.89} & $<6.48\times10^{43}$ \\
  \hline 
  AT2023vth & $2.71\times10^{43}$ (h) & 80 & $t_0 + 244.27$ & $t_0+270.72$ & {17.18} &  $<2.72\times10^{43}$ & {18.77} & $<6.30\times10^{42}$ \\  
  \hline 
  AT2024qfm & $2.71\times10^{43}$ (h) & 83 & $t_0 + 40.58$ & $t_0 + 65.53$ & {17.51} & $<2.23\times10^{44}$ & {19.23} & $<4.59\times10^{43}$\\ 
  \hline 
  AT2024wpp & $1.30\times10^{44}$ (h) & 97 & $t_0 + 356.13$ & $t_0 + 380.67$ & {17.66} & $<2.42\times10^{43}$ & {19.79} & $<3.40\times10^{42}$ \\ 

\hline
\end{tabular}
\tablecomments{(a) \citet{gutierrez_css161010} (b) \citet{perley_cow_2019}  (c) \citet{ho_koala} (d) \citet{yao_mrf} 
(e) High backgrounds and rms scatter in this light curve affect this limit.
(f) \citet{perley_xnd}
(g) TESS observations of 
AT2022tsd occurred $\sim$300-350 days after the flaring seen by \citetalias{ho_tsd_flares}.
(h) \citet{sevilla_lfbot_sample}.}
\label{tab:constraints}
\end{table*}

\subsection{Ruling out spurious late-time detections}

Given the sheer number of photometric data points from TESS and ZTF since the discovery of 
these transients,
there were several instances of formal 3-$\sigma$ detections of a source at the
location of the transient. For each such detection, we evaluated the 
probability that such detections are caused by statistical fluctuations and 
inspected the difference images at that epoch (to search for point sources at that location).

For the statistical test, we assumed that the distribution of flux residuals is
Gaussian (see Figure \ref{fig:flux_dists}), and 
calculated the probability of at least one $>3$-$\sigma$ detection occuring
throughout the span of observations after the initial LFBOT. We then used
binomial statistics to characterize the probability of observing at least one 
detection across the entire light curve, and found that none of these detections were statistically significant ($p > 0.05$). This analysis mirrors that in
\citetalias{ho_tsd_flares}. Further details about our results for the individual 
LFBOTs can be found in Appendix \ref{app:stat_tests}. We also note that the
rms scatter for TESS light curves increases considerably during times of high background,
yielding spurious detections. These were identified via visual inspection
and comparison to the local background (details on 
background estimation can be found in Appendix A of \citealt{faus_tess_ia_2023}).

\subsection{Limits on precursors}
The repeated visits of TESS to different fields of the sky allow us to search
for precursors to all 14 known LFBOTs from thousands of days before the event
to tens of days before the event. However, given
that known examples of precursors (to supernovae) can often last from tens to hundreds of days (e.g., 
\citealt{ofek_iin_precursors, strotjohann_2015, strotjohann_2021}), we focused our search
on those LFBOTs that had contiguous TESS observations prior to the initial transient
(i.e., fell within the field of view during back-to-back sectors). 
This was the case for AT2020mrf, AT2022tsd, 
AT2023fhn, AT2024aehp, and AT2026dbl (see left half of Figure 
\ref{fig:obs_timeline}). Given that there
may be evidence for significant mass loss prior to the LFBOT (e.g., \citealt{sevilla_lfbot_sample}),
precursors to these events could arise from a similar mechanism as those occurring
prior to supernovae (see, e.g., \citealt{matsumoto_sn_precursor_modeling}). However,
the luminosities of SN precursors are often $\lesssim$10\% the luminosity of the 
eventual transient (e.g., \citealt{ofek_iin_precursors})---making them difficult to detect in surveys.

For the five LFBOTs mentioned above, we searched for precursors in TESS using the
available contiguous sectors of data. We are able to rule out precursors that would have
had a luminosity comparable to that of the original transient at its peak. Such luminous
precursors are ruled out at several timescales---from tens of days before the transient
(AT2020mrf) to thousands of days before the event (AT2026dbl). Details about the precursor limits
for our LFBOTs are given in Table \ref{tab:precursor_limits}. 

\begin{table}[]
    \caption{Constraints on precursors from TESS observations of the LFBOTs prior
    to their initial detections. We only include LFBOTs that had $\geq2$ sectors of
    contiguous prior observations in TESS, a cut based on the duration of observed
    pre-supernova outbursts. ``Native'' cadence means 1800s (indicated with *), 600s (indicated
    with $\dag$), or 200s (indicated with $\ddag$), depending on when the observations occurred.}
    \centering
    \begin{tabular}{l|ll|cc}
            \hline
        \hline
        \textbf{LFBOT} & \multicolumn{2}{c}{\textbf{Observation}} & \multicolumn{2}{c}{\textbf{Limits}} \\
        & \multicolumn{1}{c}{Start} & \multicolumn{1}{c}{End} & \multicolumn{2}{c}{($10^{43}$ erg\,s$^{-1}$)} \\
        & & & (Native) & (2\,h) \\
        \hline
        AT2020mrf$^a$ & $t_0-87.75$ & $t_0 - 33.61$ & $2.35$* & $1.23$ \\
        AT2022tsd & $t_0-382.28$ & $t_0-305.53$ & $21.4$\,$\dag$ & $5.95$ \\
        AT2023fhn & $t_0-518.67$ & $t_0-466.00$ & $13.4$\,$\dag$ & $4.36$\\
        AT2024aehp & $t_0-1163.67$ & $t_0-1056.92$ & $6.82$\,$\dag$ & $2.10$\\
        AT2024aehp & $t_0-427.86$ & $t_0-378.27$ & $11.27$\,$\ddag$ & $1.53$ \\
        AT2026dbl & $t_0-1585.35$ & $t_0-1507.15$ & $7.51$\,$\dag$ & $2.22$  \\
        \hline
    \end{tabular}
    \tablecomments{(a) AT2020mrf suffered from high backgrounds, so we only used a 
    portion of the data that exhibited low scattered light as part of this analysis.}
    \label{tab:precursor_limits}
\end{table}

\subsection{Other photometric features in TESS light curves}
\label{subsec:ssos}
\begin{figure*}[h]
    \centering
    \includegraphics[width=.93\linewidth]{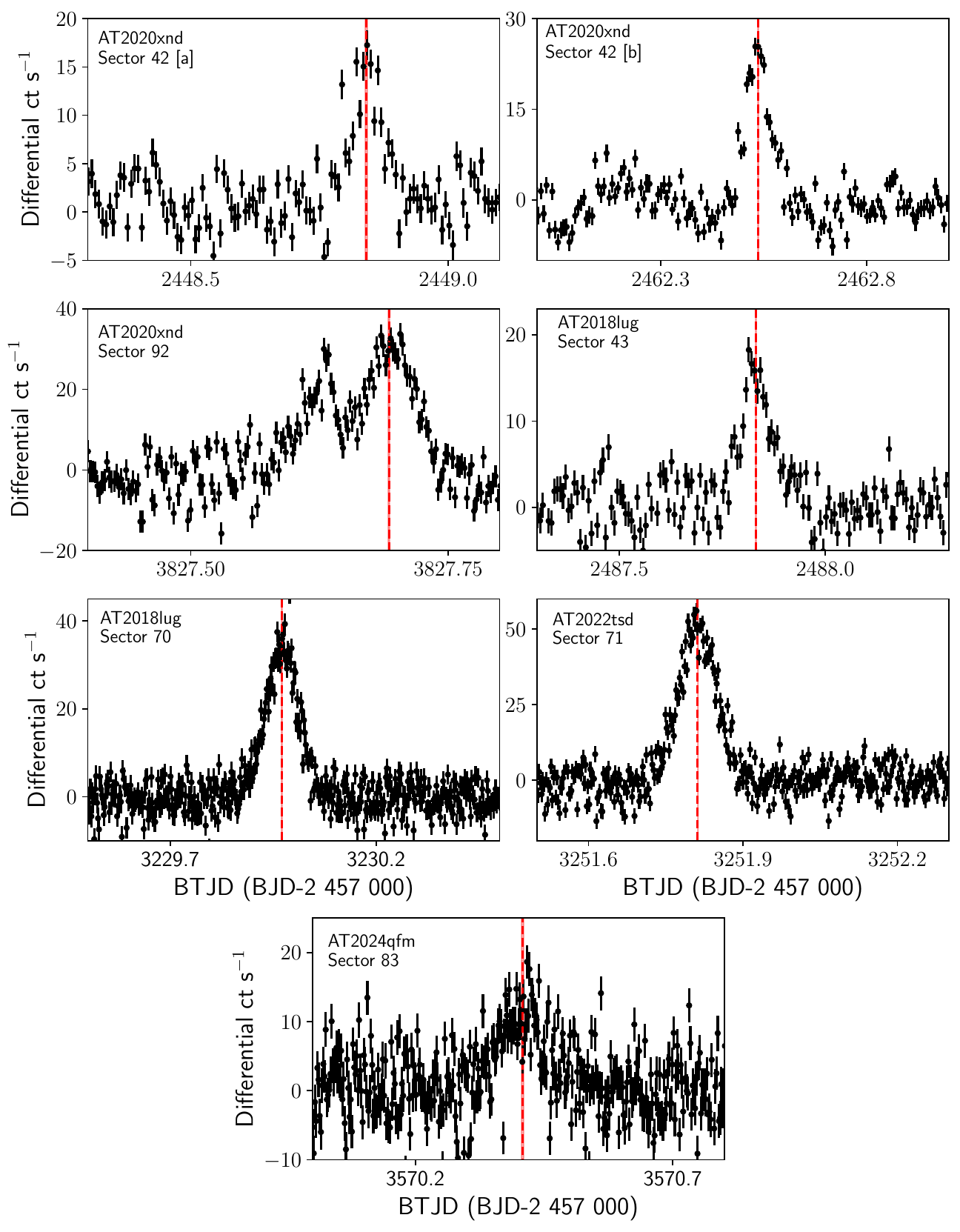}
    \caption{Flare-like features in TESS light curves. These all likely arise from SSOs (see Figures 
    \ref{fig:vetting_tool}--\ref{fig:vetting_tool_2}), which represent the 
    largest source of false positives in the search for 
    LFBOT flaring.  
    These photometric features are best fit with Gaussians in order to 
    distinguish them from stellar flares and rule this out as a possibility
    (see Table \ref{tab:ssos}). Each panel has been 
    annotated with the LFBOT and its corresponding TESS Sector.}
    \label{fig:false_positives}
\end{figure*}

Due to TESS's wide field of view and large plate scale (21''\,px$^{-1}$), there
are several sources of false positives in any analysis of transients 
(see, e.g., \citealt{tessellate,ogunwale_tequila_pipeline}). In particular, 
the largest issue affecting our search for flaring in LFBOTs was the 
presence of SSOs passing near the on-sky location of the transient.
The light curve of an SSO passing through the photometric aperture appears like 
a roughly symmetric Gaussian (e.g., \citealt{tessellate}), 
in contrast to stellar flares---which have a fast rise and a slower decay \citep{gunther_flaring}.
The seven flare-like features in the light curves---which are confirmed SSOs---are 
shown in Figure \ref{fig:false_positives}. 
We fit Gaussians to these profiles; Table \ref{tab:ssos} gives their
best-fit parameters. 

\begin{table*}
    \centering
    \caption{Results of a Gaussian fit to the flare-like profiles seen in each
    light curve. We report the LFBOT's ecliptic latitude, 
    peak time, FWHM, and the luminosity \textit{if} the flare
    arose from that particular LFBOT. The peak luminosities---if these features 
    are \textit{bona fide} flares---are higher than the peak luminosity of the
    initial transient (from Table \ref{tab:constraints}).
    AT2022tsd's flares were comparably or
    less luminous than the peak of the initial transient and were also observed in a different
    bandpass \citepalias{ho_tsd_flares}.
    Figures \ref{fig:vetting_tool}--\ref{fig:vetting_tool_2} show these targets' light
    curves, as well as cutouts from the FFIs showing SSOs passing through the aperture.}
    \begin{tabular}{cc|cllcc}
        \hline
        \hline
       \textbf{LFBOT}  & \textbf{Ecliptic Latitude} & \textbf{Sector} & \multicolumn{1}{c}{$t_{\rm peak}$} & {\bf Time since}& \textbf{FWHM} & $\nu L_\nu$\\ % FWHM = 2.355 sigma
       & (dd mm ss) & & \multicolumn{1}{c}{(BTJD)} & {\bf LFBOT} (d) & (d) & (erg\,s$^{-1}$)  \\
         \hline
        AT2020xnd & & 42 & 2448.8409(17) & $t_0 + 314.165 $& 0.071(2) & $3.97\times10^{44}$ \\ %mag 17.49
        AT2020xnd & +07$^\circ$ 01m 49.00s & 42 & 2462.5361(8) & $t_0+327.860$ & 0.075(2) & $5.77\times10^{44}$  \\ 
        AT2020xnd$^*$ &  & 92 & 3827.6928(7) & $t_0+1693.017$ & 0.045(3) & $1.41\times10^{45}$ \\
        \multicolumn{7}{c}{\dotfill} \\
        AT2018lug & +04$^\circ$ 16m 02.78s & 43 & 2487.8316(16) & $t_0+1113.919$ & 0.090(4) & $5.21\times10^{44}$ \\
        AT2018lug & & 70 & 3229.9716(6) & $t_0+1856.059$ & 0.093(2) & $1.32\times10^{45}$ \\ 
        \multicolumn{7}{c}{\dotfill} \\
        AT2022tsd & $-$09$^\circ$ 18m 47.31s& 71 & 3251.8121(5) & $t_0+421.836$ & 0.090(1) & $1.41\times10^{45}$ \\ 
        \multicolumn{7}{c}{\dotfill} \\
        AT2024qfm & +14$^\circ$ 46d 51.63s& 83 & 3570.4084(24) & $t_0+51.558$ & 0.103(7) & $2.28\times10^{44}$  \\ 
        \hline
        \hline
    \end{tabular}
    \label{tab:ssos}
    \tablecomments{* This feature exhibits a double peak, but we fit it using a 
    single Gaussian.}
\end{table*}

\begin{figure*}
    \centering
    \includegraphics[width=.98\linewidth]{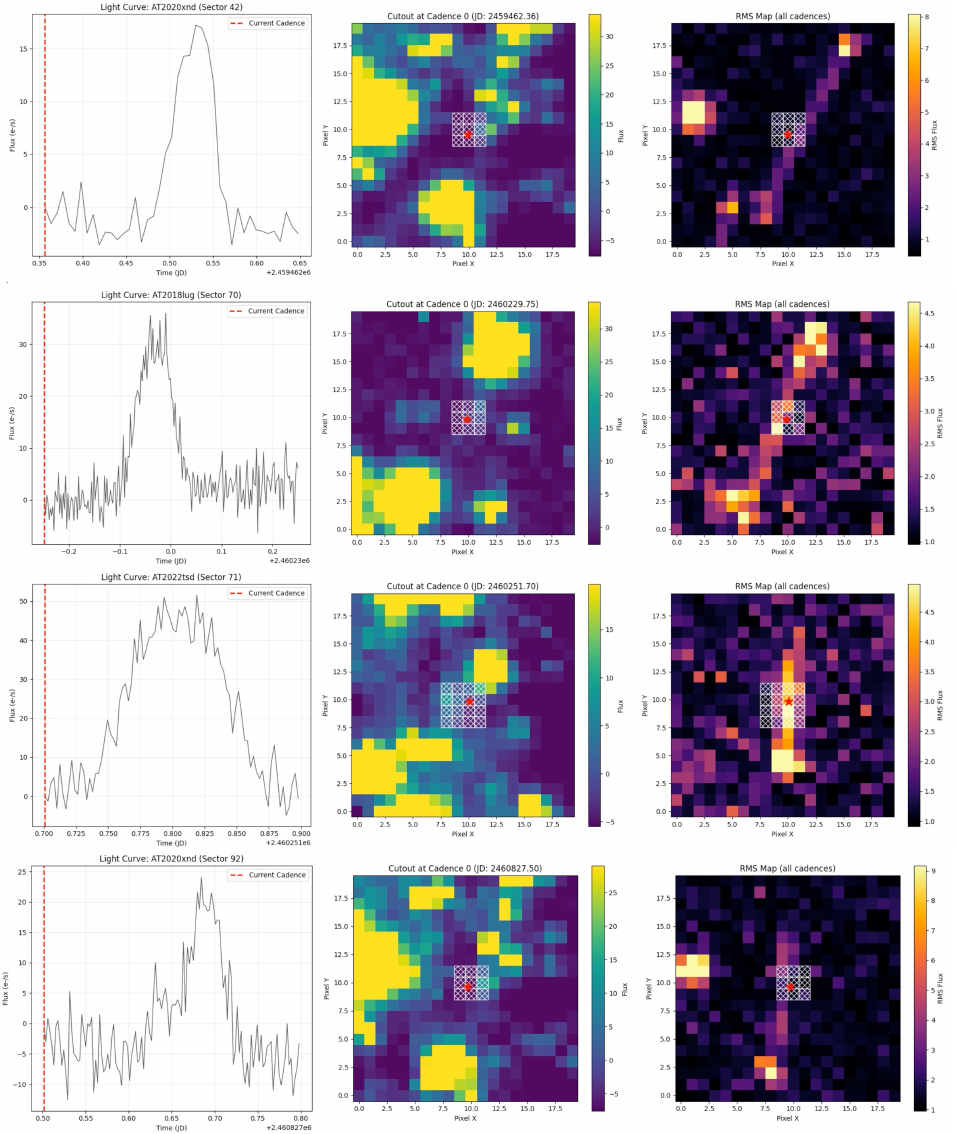}
    \caption{Screenshots of the vetting tool used to identify the passage of SSOs through a 
    photometric aperture (shown as a white cross-hatched box) centered on the LFBOT location 
    (indicated with a red star, roughly in the center of the field), for four of the seven
    LFBOTs with clear SSO signals. The dotted 
    line on the left panel moves across the light curve,
    corresponding to the cadence of the FFI in the middle panel. The right panel is a root-mean-square
    map of simple difference images created for each cadence (relative to the first image in the series);
    streaks in this RMS map also suggest that there is an object moving through the field during
    this time period. The video version of this for 
    each of the LFBOTs in Table \ref{tab:ssos} is available online. The cutouts and RMS maps have a
    spatial scale of 410''$\times$410'' (20$\times$20 TESS pixels).}
    \label{fig:vetting_tool}
\end{figure*}

\begin{figure*}
    \centering
    \includegraphics[width=\linewidth]{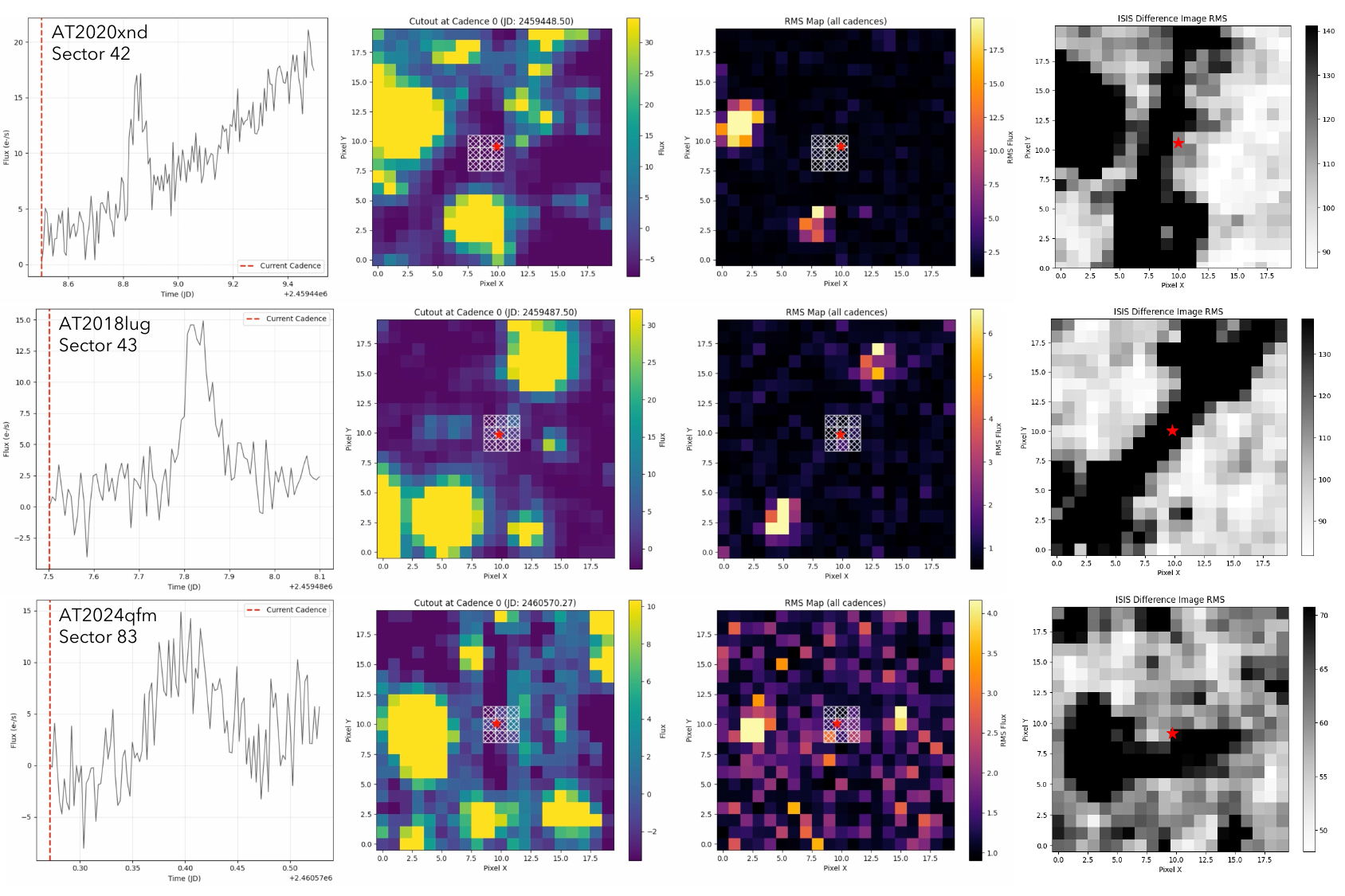}
    \caption{As Figure \ref{fig:vetting_tool}, but for those LFBOTs
    without evidence for a source moving through
    the photometric aperture in our vetting tool. We inspected the ISIS
    difference images (see Section \ref{subsec:tess_phot}) and
    constructed an RMS map from those images, and 
    there is a clear streak visible in these maps (right column). 
    These signals, on the whole, are of
    considerably lower amplitude than the corresponding signals 
    in Figure \ref{fig:vetting_tool} from the SSOs. The video version available
    online includes only the left three panels.}
    \label{fig:vetting_tool_2}
\end{figure*}

To confirm that these peaks were caused by SSOs (and 
not by flares intrinsic to the LFBOT), we used {\tt TESS-cut} \citep{tesscut} to download 
410''$\times$410'' (20 px $\times$ 20 px) cutouts around the transient from 0.5 days
before to 0.5 days after the best-fit peak time from Table \ref{tab:ssos}. We used a custom
aperture centered on the LFBOT's location to extract a light curve (with
{\tt lightkurve}; \citealt{lightkurve}), and 
created a video tracking the measured flux in the aperture alongside the FFI cutout.
Note that this light curve is different than the difference imaging light curve
described in Section \ref{subsec:tess_phot}, in order to verify that the observed
signal is visible no matter what photometric pipeline is used.
We also constructed a per-pixel RMS map, which tracks the variability in each pixel
across a given timespan. This mirrors the transient detection approach used in 
\citet{mo_gw}, and allows us to track the movement of a SSO through the image---it
will show up as a ``streak'' that is considerably higher in flux than the background.
For 4 of the 7 cases given in Table \ref{tab:ssos}, we found an object
moving through the aperture constructed using {\tt lightkurve}, as 
shown in Figure \ref{fig:vetting_tool}. For the other 3 cases, we 
did not see a clear signal in the calibrated
FFIs. As a result, we constructed an RMS map from the difference images 
and searched for evidence of SSOs therein. 
These three cases were found to have SSO ``streaks'' in the difference image RMS maps,
suggesting that the observed flares do not arise from the LFBOTs themselves. All the LFBOTs whose 
TESS light curves are shown in Figure \ref{fig:false_positives}--\ref{fig:vetting_tool_2} 
are at low ecliptic latitudes (see Table \ref{tab:ssos}). Table
\ref{tab:ssos} also shows the peak luminosity if this feature were a real
flare associated with the LFBOT. The luminosities in Table \ref{tab:ssos} are a few times the peak
luminosities of the original LFBOTs (enumerated in Table \ref{tab:constraints}).

To characterize the prevalence of false positives in TESS light curves close
to the ecliptic, we selected 500 random times during the roughly 7 years of 
TESS observations and queried the Sky Body Tracker (SkyBoT; \citealt{berthier_skybot})
to see whether there was an asteroid passing
within 80'' of that location at that time. This cone search radius is appropriate, given that
TESS's large plate scale (21''\,px$^{-1}$) can lead to contamination from objects that
other surveys, like ZTF, would consider ``far away.'' We repeated this 
experiment for the LFBOTs that had an ecliptic latitude between $-10^\circ$
and $10^\circ$ (AT2018lug, AT2020xnd, AT2022tsd, AT2023fhn, and AT2024aehp) 
and found that there was between a 15 and 20\% chance of any asteroid passing
through the aperture at any given time for each of these objects, with the
likelihood going up the closer the transient was to the ecliptic plane. 
However, the number of bright ($V<19$) asteroids passing close to the
LFBOT target was much smaller, making up $\lesssim10\%$ of SSO 
passages.\footnote{These bright SSOs are likely catalogued in the JPL Small-Body Database
(\url{https://ssd.jpl.nasa.gov/tools/sbdb_query.html}).}

To estimate the overall prevalence of detectable SSO signals, we 
make the simplifying assumption that these asteroid passages are a 
Poissonian process. We can put an upper limit on the probability of
any asteroid passage as 0.2 (from our calculated rate)
and a 95\% lower limit (from \citealt{gehrels_stats} for $n=1$)
of 0.05. The probability of seeing at least one asteroid passing
through the frame is given by $1 - e^{-\lambda t}$, where $\lambda$ is the
aforementioned rate, and $t$ is the observing duration ($\sim$25\,d). Since
not all asteroids will produce a detectable signal in TESS, we assume
that 10\% of asteroids have $V\lesssim19$ to determine which could affect
the TESS photometry. Thus, the probability
of seeing a detectable signal from a bright SSO in a 25-day long TESS observing
sector is between 5--10\%, for an observation near
the ecliptic.

\section{Discussion}
\label{sec:discussion}

\subsection{Constraints on flare luminosity and duty cycle}
Here, we constrain properties of flares from the
observed LFBOTs using limits from TESS observations, making 
the assumption that all detected ``flares'' arise from SSOs rather than from the transient. 
All upper limits are given as one-sided 3-$\sigma$ upper limits, using
Poisson statistics \citep{gehrels_stats}---this mirrors the analysis
from \citetalias{ho_tsd_flares}, who assumed a Poisson distribution
for the likelihood of detecting a flare in a given interval.

\subsubsection{Until $t_0+100$\,d} 

AT2022tsd flared from roughly 30--120\,d after the initial transient 
(see Figure 2 in \citetalias{ho_tsd_flares}). The only transient for which 
this period of time (relative to the initial detection) was observed in 
TESS was AT2024qfm ($t_0+40$--$65$\,d).
Using the photometry from \citet{sevilla_lfbot_sample}, we calculate that AT2024qfm had a 
peak luminosity of $\nu L_\nu\approx 5.7\times10^{43}$\,erg\,s$^{-1}$. 
Our results show that 
there are likely no flares with luminosities comparable to that of the original transient
during the TESS observation, using the 2\,h binned flux limit; i.e., there is no 
evidence for excess flux (above typical backgrounds/rms scatter) during the
TESS observation. At the 200\,s native cadence,
our limit rules out minutes-timescale flares that were 4$\times$ as 
luminous as the peak of the original transient. 
Our constraints on the duty cycle from the un binned light curve
is $<0.0073$ (rate of $<0.011$\,hr$^{-1}$), and from the
binned light curve, $<0.022$.

For comparison, AT2022tsd had one detected flare during the 40--65\,d window, at 
%If AT2024qfm flared identically to AT2022tsd, the flare that occurred during
%the TESS observation window (40--65 days post-explosion) would have had a luminosity of 
$\nu L_\nu \approx 5\times10^{42}$\,erg\,s$^{-1}$ in Keck $I$-band (which is slightly
narrower than the TESS bandpass, but with a similar pivot wavelength).
Such a flare would have been below TESS's sensitivity at the distance of AT2024qfm
(see Table \ref{tab:constraints}).

\subsubsection{From $t_0+100$\,d to $t_0+450$\,d} 

There are several LFBOTs, including AT2022tsd and 
a recent nearby one (AT2023vth), that were observed by TESS within roughly a year 
after their initial detections.

\paragraph{AT2022tsd} Our limit for minutes-timescale flaring using 50\,d of
continuous, 200\,s cadence observations from TESS between $t_0+380$\,d and 
$t_0+430$\,d (corresponding to $\sim$260\,d after the last
observed flares from \citetalias{ho_tsd_flares}) is $\nu L_\nu \approx 3.68\times10^{44}$\,erg\,s$^{-1}$, which
is roughly 5$\times$ as luminous as the most luminous flare seen from AT2022tsd. 
For any flaring on timescales of up to 2\,h, we have a limit
comparable to the brightest flare, of $6.7\times10^{43}$\,erg\,s$^{-1}$. 
If we assume that AT2022tsd continued to flare identically to the observations 
reported in \citetalias{ho_tsd_flares}, with the brightest flare (in $i$-band)
reaching a peak luminosity of roughly $5\times10^{43}$\,erg\,s$^{-1}$, TESS would likely not 
have detected these, even with a 2-hour bin. This weak constraint is partially
due to the distance of AT2022tsd, which is the third-furthest LFBOT we know of. 
Such an energetic AT2022tsd-like flare could have been detected for an LFBOT 
with $z<0.2$ (this would have had apparent AB magnitude of $\sim18.8$, which 
is easily detectable in a single 30-minute TESS [co-added] exposure, and marginally 
detectable in a 10-minute TESS exposure; see Figure \ref{fig:tess_limits}). 
Nevertheless, the non-detection of any flux excess even in longer bins suggests
that AT2022tsd's flares might either have stopped or have become
less luminous by the time TESS observed the field. This possibility is further
discussed in Section \ref{subsubsec:engine_shutoff}.

\paragraph{Other LFBOTs} We do not find any evidence for flaring in the
other LFBOTs at these intermediate times. 
For the more distant ($z>0.1$) LFBOTs (AT2020xnd, AT2023fhn, AT2023hkw), we find that there
is no evidence for flaring comparable to the original transient's luminosity
on timescales of 2\,h or shorter. For minutes-duration flaring, our limits
are above the original transient's luminosity. We do note that the second 
flare observed from AT2022tsd (at roughly $t_0+28$\,d) exhibited a luminosity 
that was comparable to, if not greater than, the peak luminosity of the
early LFBOT light curve. 

For the nearby ($z<0.1$) LFBOTs (AT2023vth and AT2024wpp), our limits are considerably more stringent.
At the native 200\,s cadence, we can rule out flaring comparably as luminous as the original
transient for AT2023vth, and {\it flaring that over 5$\times$ as faint as
the peak of AT2024wpp} out to a year after the initial observation. Binning to 2\,h
strengthens these limits (see Table \ref{tab:constraints}). This last non-detection suggests either that
the flaring turns off after a year (as was seen in AT2022tsd) or that there is
another phenomenon at play (e.g., beaming) that has prevented us from observing
flaring in other events apart from AT2022tsd
(see Section \ref{subsec:physical_expl} for further discussion). Our limits from TESS
(for a 2-hour bin) are comparable to the limits for early-time flaring from 
\citet{ofek_wpp_limits}, of roughly $3\times10^{42}$\,erg\,s$^{-1}$.

\subsubsection{Late times: $t_0+450$\,d and beyond} 

There are several LFBOTs in our
sample that have observations over a year to several years after the initial 
transient. We have demonstrated that AT2022tsd likely stops flaring with a comparable
luminosity to the initial flares after roughly a year, and can extend this limit 
to other events. For the farther LFBOTs with late-time observations
(AT2022abfc, AT2020mrf, AT2018lug, AT2020xnd), our upper limits on years-long timescales are
within an order of magnitude of the initial luminosity.
For the nearby LFBOTs with late-time observations (AT2018cow and CSS161010), we 
can constrain the luminosity of any potential extremely late-time flaring to be 
a few hundred times as faint as the initial transient, suggesting that they are 
not as active at late times after a potentially stable accretion disk has formed
\citep{chen_hst_uv_cow_ii,inkenhaag_hst_uv_cow}.

Our TESS observations of AT2018cow coincide with Hubble Space Telescope observations of this
source, reported in \citet{sun_hst_uv_cow}. In the F814W bandpass (which overlaps much of the
TESS bandpass), \citeauthor{sun_hst_uv_cow} report a detection of a source at the position 
of AT2018cow with a magnitude of $25.82\pm0.19$ (Vega). To within a small correction term
(for Vega to AB conversion), this source is roughly 800$\times$ fainter than the
TESS detection limit. \citet{inkenhaag_hst_uv_cow} theorize that this source is 
powered by accretion, with a persistent luminosity that is roughly 1000$\times$ fainter
than the detection limit of TESS during this time. 

% A flare detectable by TESS from this putative source (this flare would have to have a
% luminosity of $6.88\times10^{40}$\,erg\,s$^{-1}$ or greater) from a persistently
% accreting source as faint as AT2018cow would be unphysical.

\subsubsection{Flaring duty cycle and observability}

Our observations suggest that there is a point at which the flaring ``switches off''
or becomes less luminous
for AT2022tsd, and suggest that there is no evidence for flaring beyond 100\,d in any of the
other transients (as well as no flaring at early times in AT2024qfm). These results
allow us to constrain the duty cycle of these flares when TESS was
observing them. Our approach mirrors that of \citet{ofek_wpp_limits}, and we
use the formula $\Delta\,t/(60\cdot T_{\rm obs})$ for the duty cycle. Here, $\Delta\,t$ 
represents the timescale over which we expect the flares to be active, and $T_{\rm obs}$
represents the observation duration. Since the Poisson 
single-sided upper limit for a 3-$\sigma$ detection (99.8\% confidence) 
is 6.6 (from \citealt{gehrels_stats}), the upper limit on the 
duty cycle is $<6.6\Delta\,t/60T_{\rm obs}$. 

If we assume an average flaring 
duration of $\sim$40\,min (based on AT2022tsd)
and an average TESS observation duration of 20--25\,d, 
then the upper limit on the duty cycle is $<0.007-0.01$ during the times of 
observation. Relaxing the upper limit 
to 2-$\sigma$ yields a duty cycle of $<0.003-0.004$. 
For continuous observations with two 25\,d sectors (as was the case for 
AT2018cow, AT2018lug, AT2020mrf, and AT2022tsd), or the case of longer sectors (e.g., 
Sectors 97 and 98---the case for AT2024wpp, CS161010, and AT2022abfc),
the $T_{\rm obs}$ increases to roughly 55\,d. This larger $T_{\rm obs}$ strengthens
the constraint on the duty cycle to $\lesssim0.004$ (3-$\sigma$), under
the assumption that any flaring occurring in the transient would have been 
luminous enough for TESS to detect. The duty cycle limit
and observation time are inversely related; these limits are reduced by 
factors of $1/N$ for longer observations, 
where $N$ is the number of sectors during which the transient was observed. 
These upper limits are comparable to, if not more stringent than, the
estimates for the duty cycle of the flaring of AT2022tsd (between 0.02--0.5; 
see Extended Data Table 3 in \citetalias{ho_tsd_flares}).

\begin{figure*}
    \centering
    \includegraphics[width=\linewidth]{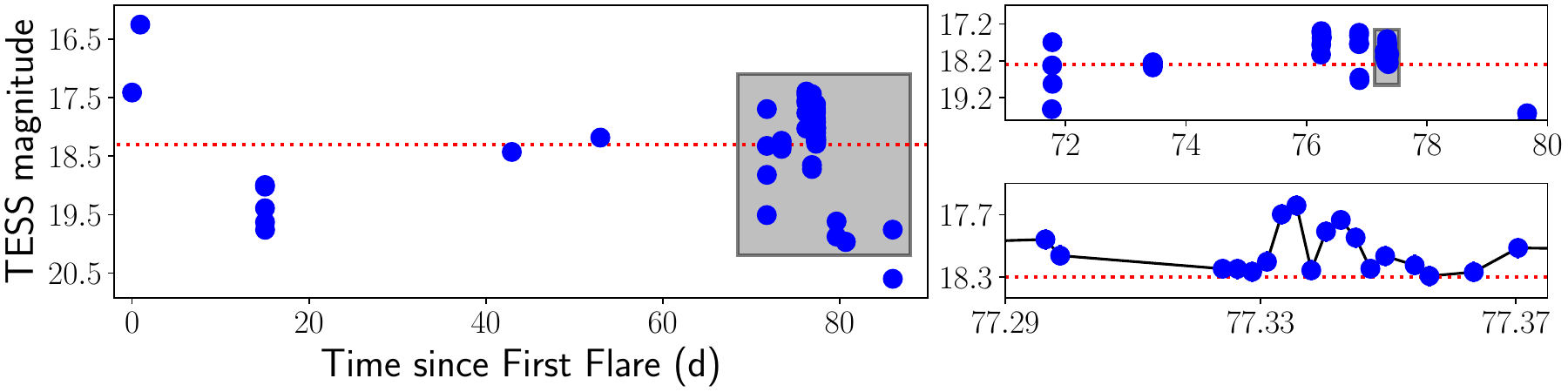}
    \caption{A simulated light curve of AT2022tsd-like flares in
    the TESS passband, situated at $z=0.08$. TESS 
    has the necessary temporal resolution to individually resolve several distinct
    flaring episodes, as shown in all three panels. The red dotted line 
    in all panels represents the theoretical lower limit for TESS detections at the
    native cadence of 200\,s (see Figure \ref{fig:tess_limits}). However, only three known LFBOTs lie within 
    this redshift---the first two discovered (CSS161010 and AT2018cow; see Table \ref{tab:fbot_info}), 
    and AT2023vth \citep{sevilla_lfbot_sample}. Within this volume, there should
    be 0.2--2\,LFBOTs yr$^{-1}$, based on the rates from \citet{perley_wpp}. The 
    $k$-correction used in this analysis assumes an underlying source
    SED of $F_\nu\propto\nu^{-1.6}$ \citepalias{ho_tsd_flares}.}
    \label{fig:tess_flares}
\end{figure*}

We simulated how AT2022tsd-like flaring would appear in TESS if an LFBOT
were situated at $z = 0.08$, which is further than AT2018cow and CSS161010 
but closer than AT2022tsd. These flares peak at
$T_{\rm mag}\sim16.5$, with the substructure clearly resolved at TESS's 200\,s
cadence (shown in Figure \ref{fig:tess_flares}). While we do
not yet have a robust constraint on the rate of late-time flaring in LFBOTs, 
we estimate the detection rates of these flares using current and
upcoming facilities in Table \ref{tab:survey_rates}. These calculations are
further discussed in Section \ref{subsec:ground_surveys}.

\subsection{Physical Implications}
\label{subsec:physical_expl}

Our non-detection of confirmed late-time flaring in any of the other LFBOTs (apart from AT2022tsd) 
across several timescales has implications for the duration of central engine activity 
and the beaming of observed emission. We consider both hypotheses separately and
characterize each one's effects on flare detection.

\subsubsection{Decline or ``shutoff'' in engine power}
\label{subsubsec:engine_shutoff}
First, our results suggest that the engine power declines or shuts off on a 
timescale of hundreds of days, given the non-detection of flaring
after 100\,d in AT2022tsd. The redshift of AT2022tsd and the sensitivity of our 
limits from TESS mean that we cannot constrain the rate at which the engine power 
declined. If the engine is fallback accretion from stellar collapse 
(e.g., \citealt{quataert_kasen_swift1644,quataert_bh_accretion_stellar_collapse, chrimes_lfbot_ccsne}), 
the engine power drops precipitously after much of the mass is accreted. 
The relevant timescale for this phenomenon 
is the free-fall time $t_{\mathrm{ff}}$ \citep{quataert_kasen_swift1644}:
\begin{equation}
    t_{\rm ff}(r) = \frac{\pi r^{\frac{3}{2}}}{(2GM)^{\frac{1}{2}}} \approx 702 \left(\frac{r}{10^{14}\,{\rm cm}}\right)^{\frac{3}{2}} \left(\frac{M(r)}{10\,M_\odot}\right)^{-\frac{1}{2}}\,{\rm d},
\end{equation}

\noindent where $r$ and $M$ are the initial radius and the mass of the infalling 
material, respectively. Thus, in this scenario a timescale of hundreds of days would 
require an extended ($\sim10^{14}\,$cm) stellar progenitor, like a red supergiant.

The calculations in \citet{metzger_fbot_theory_2022} suggest that 
the luminosity available to power accretion by the central engine drops as
roughly $t^{-2.1}$, as observed in light curves of AT2018cow and other LFBOTs. 
The flaring phenomenon may be related to transitions in accretion state,
as suggested in \citet{migliori_cow} for AT2018cow in order to explain the
late-time X-ray observations, though flaring was not detected in this event 
(despite intensive multi-wavelength monitoring).

\subsubsection{Beaming of the flare emission}
The red color of the flares in AT2022tsd suggests that this emission is
nonthermal (possibly of synchrotron origin); together with the high 
brightness temperature, this suggests it is somewhat beamed. 
This would require the observer to be on-axis to observe
these flares. \citetalias{ho_tsd_flares} estimated the outflow opening
angle of AT2022tsd to be roughly 30$^\circ$ based on the non-detection
of flaring from the other 6 (at the time) known LFBOTs. From our 
constraints on the newer LFBOTs, we can constrain
the beaming fraction $f_b$ to be between $\frac{1}{14}$ and $\frac{1}{6}$,
where the denominator is the number of LFBOTs we have detected so far 
(14 as of this work; 6 in \citetalias{ho_tsd_flares}). 
Since $f_b = 1-\cos\theta$, where $\theta$ is the outflow opening angle,
we find that a beamed outflow could be weakly collimated, with an estimated 
beaming angle of 5--30$^\circ$. This uncertainty calculation uses
the 90\% confidence lower limit for Poisson statistics 
(from \citealt{gehrels_stats}), 0.0513 for $n=1$.

From this simple beaming angle calculation (which has a high uncertainty, given
only one detection of LFBOT flaring, and also does not take into account
any potential flare ``shutoff'' timescale), it appears that
LFBOT outflows could be wider than those observed in gamma-ray bursts
(see, e.g., the jet opening angle distributions for long GRBs from 
\citealt{goldstein_lgrb_opening_angles}). The presence of a collimated
outflow could also explain the asphericity observed in certain LFBOTs 
(e.g., \citealt{maund_cow_polarization}), and further support the hypothesis
that while AT2018cow was well-monitored, we were not observing it
at an appropriate orientation (i.e., pole-on) that would enable us to observe 
a launched jet and any concomitant late-time flaring. These results are consistent 
with inferences on the jet geometry from \citet{margutti_cow_2019}, who
find that the radio observations of AT2018cow can be explained by either
low-energy, highly-collimated jets \textit{or} high-energy, 
weakly-collimated jets propagating into a wind medium with low mass 
loss rates ($\dot{M} < 10^{-4}$\,M$_\odot$\,yr$^{-1}$). As
the ejecta decelerates, for an off-axis relativistic jet, 
we theoretically could observe a ``jet break''-like phenomenon 
(as seen in GRBs), though the late-time millimeter and radio 
observations of AT2024wpp are difficult to explain using such a model \citep{nayana_wpp}.

\subsection{TESS and LFBOTs}
Our results demonstrate that, for a nearby LFBOT such as AT2018cow, CSS161010, or AT2023vth
($z\lesssim0.08$) that exhibits flares with luminosities comparable to the original transient
(approximately $10^{43}$\,erg\,s$^{-1}$), TESS should be able to confidently detect them
and capture temporal sub-structure, especially at its current 200\,s cadence. 
If the flares in AT2022tsd were at the 
distance of AT2018cow, they would have a peak magnitude 
$T_{\rm mag}\sim12$--15---comfortably detectable by TESS.

Assuming a fiducial limiting (AB) magnitude for 
TESS of $\sim18$ at its 200\,s cadence, and an LFBOT peak magnitude
range of $-22\leq M \leq -20$, TESS could detect the peaks of 
LFBOTs out to a luminosity distance of $\sim$1\,Gpc ($z\sim0.2$)
for the brightest events. However, a single detection (or several)
at or around the LFBOT peak would not take advantage of TESS's key
advantage, its 200\,s cadence. Realistically, TESS could provide a 
well-sampled light curve for an LFBOT within a few hundred Mpc ($z\lesssim0.1$;
see, e.g., Figure \ref{fig:tess_flares}).
TESS light curves could be binned to longer intervals (e.g., a few hours)
to help characterize the rise to peak and the decay, expanding the horizon
for LFBOT detectability, though this is contingent on the level of the
background from scattered light due to Earth- and/or Moonshine entering the
TESS camera. High-cadence observations of the initial transient could
also help us study its evolution on short timescales (minutes to hours).

While the duration of the most intense episode of flaring observed
in \citetalias{ho_tsd_flares} was roughly 40\,min, we binned 
the light curves to 120\,min in order to establish deeper limits 
(by a factor of $\sim$5--10) than possible using the 200\,s data alone. 
Even if the flare spanned two of these 120\,min
bins, the flux excess would have been detectable at a 3--5-$\sigma$ 
level, especially for the nearby LFBOTs. However, this binning
approach obscures any sub-structure in a flare. This issue could
be overcome for a nearby LFBOT by obtaining TESS short-cadence
observations (at either 2 minutes or 20 seconds)\footnote{TESS obtains
high-cadence observations of a 10\,px $\times$ 10\,px
region of the sky around $\sim$10\,000 (community-selected) targets
at 2\,min cadence and $\sim$3\,000 targets at 20\,s cadence.} in the months after the initial
transient, though the likelihood of one happening as nearby as AT2018cow
or CSS161010 is quite low.

\begin{figure}
    \centering
    \includegraphics[width=\linewidth]{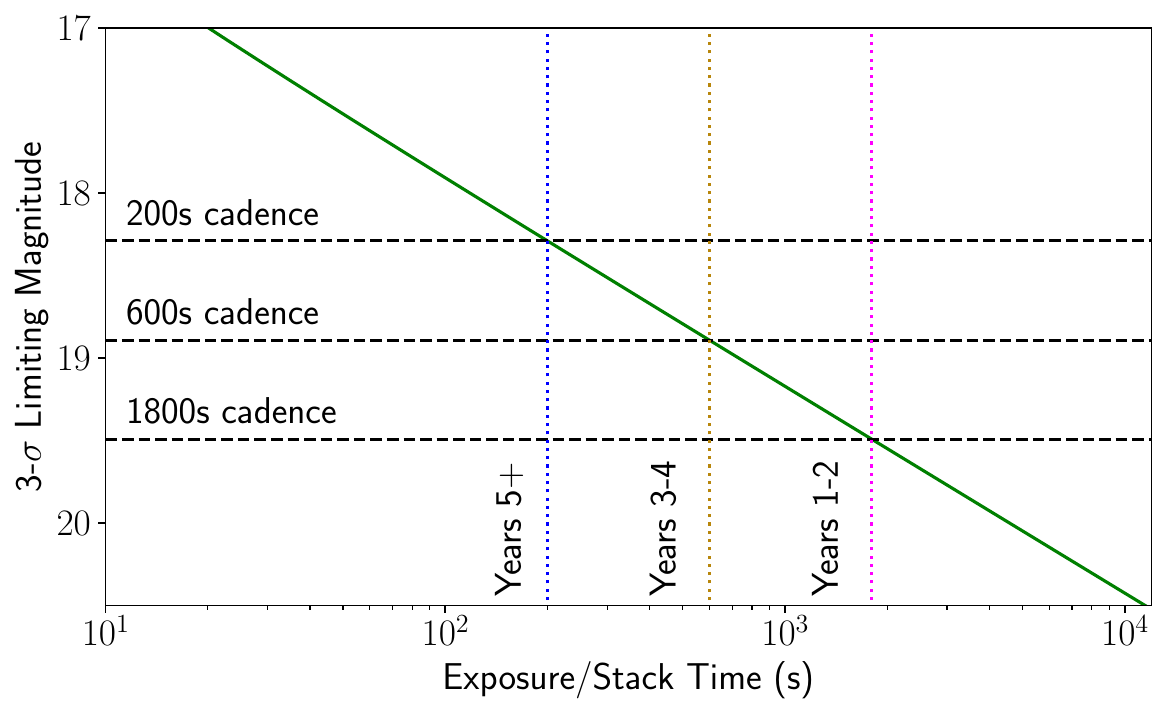}
    \caption{The theoretical TESS limiting magnitude as a function of 
    exposure time (as calculated from the CCD equation) is 
    shown in green, with the various observational cadences throughout
    the mission duration indicated. As inputs to the CCD equation, we assumed
    a dark current of 0, a fiducial quantum efficiency of 0.95\footnote{This has been
    found to vary across the bandpass; see, e.g., Figures 9--10 in
    \citet{krishnamurthy_tess_detector}.}, 
    a read noise of $10$\,e$^-$\,px$^{-1}$\footnote{Sources: Figure
    4.5, \S 6.2, and \S 6.1 in the TESS Instrument Handbook, respectively, for
    the three values:
    \url{https://archive.stsci.edu/files/live/sites/mast/files/home/missions-and-data/active-missions/tess/_documents/TESS_Instrument_Handbook_v0.1.pdf}},
    an aperture size of 4 px, an effective exposure time\footnote{20\% of the sub-exposures are clipped
    due to onboard cosmic-ray mitigation, reducing the
    effective exposure time from the nominal cadence (see the Appendix of 
    \citealt{jayaraman_grb_2024})} of 160\,s, and
    an average sky background of 100\,e$^-$\,px$^{-1}$\,s$^{-1}$. The sky background 
    was empirically calculated using 
    several FFIs from different sectors with low scattered light background.}
    \label{fig:tess_limits}
\end{figure}

Figure \ref{fig:tess_limits} shows the theoretical 3-$\sigma$ detection
limits for TESS as a function of exposure time for an assumed set of 
input values to the CCD equation,
\begin{equation}
    {\text{S/N}} = \frac{S_{\rm obj} \cdot \sqrt{t \cdot QE}}{\sqrt{S_{\rm obj} + n_{\rm pix} \left(S_{\rm sky} + \frac{S_{\rm dark}}{QE} + \frac{R^2}{QE \cdot t}\right)}},
\end{equation}
where $S_{\rm obj}$ is the flux of the object (e$^-$\,s$^{-1}$), $t$ is the
effective exposure time, QE is the quantum efficiency, $n_{\rm pix}$ is the
number of pixels in the photometric aperture, $S_{\rm sky}$ is the sky 
background (also in e$^-$\,s$^{-1}$), $S_{\rm dark}$ is the dark current,
and $R$ is the read noise. Note that at the flight temperatures of the
TESS detectors ($\sim$200\,K), dark current has been found to be negligible during lab testing
\citep{thayer_dark_current}.
Other input parameters are enumerated in the caption to Figure \ref{fig:tess_limits}.

One other crucial consideration when studying short-timescale phenomena with TESS is its
onboard cosmic-ray mitigation algorithm, which clips roughly 20\% of the 
exposures when it co-adds sub-exposures to create the ``final'' FFI. 
For phenomena varying on timescales of a few seconds, this onboard
technique could clip a large fraction of astrophysical flux from the source. This effect was 
investigated by \citet{jayaraman_grb_2024}, who found that in the worst cases, 
TESS's onboard CRM could clip over 50\% of the flux from a source varying on timescales of 
a few seconds. LFBOT flares were found (in \citetalias{ho_tsd_flares})
to vary on timescales of $\sim$30--60\,s, which could be clipped as part 
of TESS's co-addition when it constructs the FFIs.
This would thus lead to a systematic underestimate of the flux from a rapidly-varying 
source. Contemporaneous observations from another observatory can help diagnose
this issue; \citet{jayaraman_grb_2024} found that flux estimates 
would need to be revised upward by 20--25\% for a more accurate value, when compared 
to high-energy prompt light curves for GRBs.

\paragraph{Rates} \cite{ho_fbot_ztf_2023} initially calculated
the volumetric rate of LFBOT events
to be 70\,yr$^{-1}$\,Gpc$^{-3}$; this estimate was revised downward by 
\citet{perley_wpp} to be between roughly 1--10\,yr$^{-1}$\,Gpc$^{-1}$, with
the rate of bright 2024wpp-like events being even lower---with an upper
limit of 0.7\,yr$^{-1}$\,Gpc$^{-1}$. In a simplistic analysis (i.e., not accounting
for extinction, assuming a 100\% observing efficiency, and achieving full sky coverage
from TESS every $\sim$2\,yr), we find that TESS, which covers $\sim5\%$ of the sky 
every Sector, should be able to detect the initial emission of up to one LFBOT
per year. Such a detection has not yet been achieved, as there has not yet 
been an LFBOT (discovered by another survey) that has fallen within 
TESS's field of view at that time. 
Untargeted searches (e.g., \citealt{tessellate}) in 
archival TESS images may yield such transients, as occurred during the Kepler
mission for KSN 2015K \citep{ksn_2015k}, referred to at the time as a 
``fast-evolving luminous transient.'' However, the rates for such untargeted discoveries
are highly uncertain: Table 2 in \citeauthor{tessellate} suggests that there are
$1093\pm1092$ ``FBOTs,'' with peak absolute magnitudes from $-16$ to $-22$,\footnote{The
rate of LFBOTs, i.e., AT2018cow-like transients, is likely a small fraction of this number. 
However, there are likely 
several LFBOTs in southerly fields (inaccessible to ZTF) that could be identified through
an untargeted search in TESS. The limiting magnitude for such a search is rather bright, though,
with 50\% of injected sources at 16th magnitude being recovered by the {\tt TESSELLATE} detection 
pipeline \citep{tessellate}.}
that could be recovered from the first $\sim$6\,yr of TESS observations.

\subsection{LFBOT flaring from high-cadence surveys}
\label{subsec:ground_surveys}

\begin{table*}
    \caption{Properties of different high-cadence and their efficacies in detecting late-time flaring in LFBOTs. Further
    information about LAST can be found in \citet{ofek_last} and \citet{konno_last_transient_detection_pipeline}, LSST/Rubin in 
    \citet{ivezic_lsst}, ULTRASAT in \citet{ultrasat_i} and \citet{ultrasat_ii}, and the Argus Array in \citet{argus_array}. The calculations
    in this table rely on the following
    assumptions: (a) the luminosity of each flare is 10$^{43}$\,erg\,s$^{-1}$ in $r$-band, 
    (b) the spectral index of these flares is 
    $F_\nu\propto\nu^{-1.6}$ (as reported in \citetalias{ho_tsd_flares}), and (c) there are 100 episodes of flaring, each lasting $\sim$25\,min,
    occurring over 90\,d (similar properties as AT2022tsd). If we assume that all LFBOTs flare in 
    such a manner, then we can use the rate of $1$--$10$\,Gpc$^{-3}$\,yr$^{-1}$ \citep{perley_wpp} for the calculations of flaring detectability.}
    \centering
    \begin{tabular}{c|crccccc}
    \hline\hline
     {\bf Survey}  & {\bf Exposure} & {\bf Cadence} & {\bf Field of} & {\bf Limiting} & {\bf Maximum} & {\bf Volume} & {\bf Rate}  \\
      & (s) & (min) & {\bf View} (deg$^2$) & {\bf Magnitude} & $z$ & (Gpc$^3$) & (yr$^{-1}$) \\
     \hline
       TESS & 200 & 3.33 & 2\,304 & 18.2 & 0.08 & 0.007 & 5.5--54 \\
        % TESS (\textit{binned}) & 7\,200 & '' & 20.2 & 0.43 & 3.44 & 3.4--34 \\
       LAST & 20 ($\times20$)$^a$ & $\lesssim60^b$& 355 & 21.0 & 0.19 & 0.015 & 0.5--5 \\
       ULTRASAT & 300 & 5$^c$ & 204 & 22.5 & 0.23 & 0.014 & 6.9--69\\
       LSST DDF ($r$-band) & 30 & $\lesssim500^d$ & 9.6 & 24.7 & 0.80 & 0.10 & 0.9--8.6 \\
       Argus Array & 60 & $1^e$ & 8\,000 & 20.5 & 0.17 & 0.97 & 8--80\\
       \hline
    \end{tabular}
    \tablecomments{(a) In the slow mode, LAST scans
    the sky in 20$\times$20 s exposure visits; in the fast mode, the visits are reduced to 5$\times$20\,s 
    exposures. (b) Based on the fast cadence survey (see Figure 5 in \citealt{ofek_last}, which assumes 300\,s exposures). (c) This is for
    the high-cadence survey (21\,hr\,d$^{-1}$); the low-cadence survey (3\,hr\,d$^{-1}$; 6800\,deg$^2$) will have
    a much lower detection rate. (d) It is anticipated that DDF fields will get 10$\times$ as many visits as
    the Wide, Fast, Deep survey fields, which will be observed every $\sim$3\,d. Further information about the WFD
    and DDF surveys can be found at \url{survey-strategy.lsst.io} (e) Argus is a 
    ``drift scanning'' survey, which tracks the sky as the Earth rotates.}
    \label{tab:survey_rates}
\end{table*}

We evaluate the ability of high-cadence (minutes-to-hours) surveys---LAST, 
ZTF, the Legacy Survey of Space and Time (LSST)
at the Vera Rubin Observatory, the Argus Array, and the upcoming 
ULTRASAT satellite---to detect late-time flaring in
LFBOTs. We assume that the most luminous flares have a 
luminosity of 10$^{43}$\,erg\,s$^{-1}$
and use this value in the calculations underlying 
Table \ref{tab:survey_rates}. We convert this to a flux and calculate
the redshift at which we could detect this flux; we correct the volume
enclosed at this redshift for a given survey's field of view to find the
volume probed. We assume that there are 100 flares, each lasting 25 minutes over a period of 90 days
(\citetalias{ho_tsd_flares} detected 12 flares from AT2022tsd, but there may 
have been more that were not observed).
We also assume that the flare emission has an SED of $F_\nu\propto\nu^{-1.6}$,
identital to AT2022tsd.
We then take the peak luminosity
of the flare across a variety of redshifts (taking into account 
$k$-corrections and corrections for the comoving volume) and the 
probability of detecting a flare, given a survey cadence. For Argus 
specifically, given that it is not a slewing survey, rather observing the 
sky as the Earth rotates, we take into account the accessible range of
the sky and how long a source transits over Argus (i.e., 
remains visible at its declination). We find that the highest
rates of flare detections will arise from the Argus Array, ULTRASAT, and TESS.

In calculating the rates of flaring detectability shown in Table \ref{tab:survey_rates}, we assume that
every LFBOT flares like AT2022tsd and so adopt the LFBOT rate. In reality, 
we have shown that not every LFBOT is observed to flare in an AT2022tsd-like 
manner. If not every LFBOT flares, then the calculated rates should be
scaled by the fraction of LFBOTs empirically observed to flare---1/14
in the most pessimistic case, based on this work and \citetalias{ho_tsd_flares}. 
This multiplicative factor
could also account for beaming, i.e., where we are observing the 
majority of LFBOTs off-axis, and only saw AT2022tsd on axis (and hence
observed flaring). We also note that for LFBOTs flaring at higher redshifts, 
given the $F_\nu\propto\nu^{-1.6}$ dependence, much of the emission should be in the 
NIR, though this is a difficult band to observe with ground-based observatories,
given the high sky backgrounds. We thus focus on $r$-band observations
from ground-based surveys to capture flares from nearby LFBOTs in this analysis.

\paragraph{LAST} This survey telescope was already used to search for
flares from AT2024wpp in \citet{ofek_wpp_limits} and will be fully 
operational by the end of 2026. A LAST ``node'' is comprised 
of 48 28-cm telescopes equipped with CMOS detectors 
that provide a total field of view of 355\,deg$^2$
and a collecting area equivalent to a 1.9-m telescope. The 5-$\sigma$ limiting 
magnitude (from \citealt{konno_last_transient_detection_pipeline})
for the $20\times20$\,s exposures is 20.6$\pm$0.4.

In our rate calculations, we assume LAST's fast-cadence survey strategy
(as discussed in \citealt{ofek_last}), which scans the sky up to 
8 times in a given night (in areas below airmass 2). In this case, we
obtain a rate of flare detection
of 0.5\,yr$^{-1}$. However, this number comes with caveats; in particular, 
LAST can be used for \textit{targeted} follow-up of LFBOTs detected by 
other surveys in order to search for flares, as in \citet{ofek_wpp_limits}. 
This will likely increase
the rates of flaring that would be detected, or provide further, more stringent
constraints on the timescale and luminosity of flaring in LFBOTs. It could also 
be used to search for flaring in other transients, as this is an unexplored
regime.

\paragraph{ZTF}
\citetalias{ho_tsd_flares} set upper limits on flaring in 6 of the first 7
LFBOTs discovered (not including AT2022tsd) using ZTF. For the remaining 6
LFBOTs, we do not find any detections consistent with flaring behavior 
(on minutes-duration timescales) in the months to years after the transient.
ZTF's uneven sampling and 1-day
cadence observations make it difficult to calculate upper limits on a duty
cycle, in contrast with the limits we are able to set using TESS (and those
set with LAST in \citealt{ofek_wpp_limits}). We discuss spurious ZTF
detections at the locations of these transients in Appendix \ref{app:stat_tests} and
calculate the chance occurrence probabilities of such detections.

Given ZTF's observing strategy, it will be able to detect epochs at which
flaring is occurring (as in \citetalias{ho_tsd_flares}), but cannot 
resolve the minutes-timescale duration of this behavior. Its ongoing 
nightly-cadence survey in $gri$ bands of the region of the sky overlapping
Rubin's footprint will help us detect any potential flux excesses (i.e., 
deviating from the typical LFBOT decay) that could herald flaring behavior.
In addition, its higher cadence (compared to Rubin's WFD survey) will allow a better
constraint on the peak time of the initial LFBOT.

\paragraph{LSST/Rubin}
While the cadence of LSST/Rubin's wide-fast-deep (WFD) survey will be insufficient
to study such flares, LFBOTs found in the deep drilling fields (DDFs; spanning
70\,deg$^2$) may be
amenable to flare identification, given the sub-day cadence observations. 
However, these fields represent an extremely small
portion of the overall LSST survey, and given the low (and highly uncertain)
rate of LFBOTs, it is difficult to determine whether the DDF cadence and 
observations will yield any useful detections of LFBOTs or flaring. Our predictions 
in Table \ref{tab:survey_rates} suggest that the DDFs will yield, in an
optimistic case, around 9 LFBOT flares per year. However, the WFD will still be
useful as a discovery engine for LFBOTs themselves---the three-day cadence
is enough to track their early evolution and help constrain 
the blackbody parameters. Moreover, the WFD will enable the identification
of LFBOTs at higher redshifts, across cosmic time. Our preliminary estimates
suggest that Rubin could detect tens of LFBOTs throughout its survey (though
a far smaller number of flares). Complementary
higher-cadence surveys whose fields of view overlap with 
Rubin (LAST, Argus) can then be used to search for flaring in these events. 

\paragraph{Argus Array}
The Argus Array \citep{argus_array}, an upcoming survey telescope with a field of 
view of 8\,000\,deg$^2$ that will be able to image the sky at a 60\,s cadence
with a limiting magnitude
of $\sim20-20.5$, represents one of the highest-cadence ground-based transient 
surveys to date. Assuming flaring behavior comparable to that of AT2022tsd,
Argus could detect up to 80 flares in a year---making it among the most promising 
surveys of the ones in Table \ref{tab:survey_rates}.
We do note that to resolve minutes-duration flares (rather than simply observe
the peaks of several flaring episodes), the volume that Argus can probe 
reduces somewhat, and the rates that we estimate in Table \ref{tab:survey_rates}
will reduce concomitantly. Regardless, it still retains its considerable advantages over
other surveys with its large field of view and high cadence.

\paragraph{ULTRASAT}
This satellite will conduct a survey of the sky at near-UV wavelengths
\citep{ultrasat_ii}, with its high-cadence survey observing roughly 
200\,deg$^2$ near the ecliptic poles at a 300\,s cadence for six months
at a time. With a limiting magnitude of $\sim$22.5, this will be sensitive
to both the initial blue emission of LFBOTs, as well as their flares (which
are anticipated to be fainter at blue/UV wavelengths due to the steep SED
observed in \citetalias{ho_tsd_flares}). While the 5-minute cadence will 
obscure some of the sub-structure in the flares, the spatial association 
with an LFBOT and contemporaneous observations planned using 
LAST\footnote{\url{https://www.weizmann.ac.il/ultrasat/science-mission/modes-of-operation/modes-of-operation}} (in a redder band) will allow us to rapidly characterize these flares and 
trigger other facilities to observe and monitor these events, perhaps at 
a higher cadence. We anticipate that
ULTRASAT will, alongside Argus, be the facility that can
identify the most LFBOT flares throughout its lifetime.

\section{Conclusions}
In this paper, we have presented TESS observations of LFBOTs
to constrain the presence and incidence of late-time optical flaring
in 12 of the 14 known such events. We detected seven flares in total from 
four unique LFBOTs. All of the seven were found to arise from SSOs passing
through the photometric aperture.

On the timescales over which flaring was observed in AT2022tsd, we do not
observe any confirmed flaring comparable to the original luminosity in the LFBOT 
AT2024qfm, that clearly arises from the transient itself. 
For intermediate timescales (between 100--450\,d after the initial
transient), we can rule out flaring that is roughly 10 times as faint as the original
event for two nearby LFBOTs ($z\lesssim0.1$). At late timescales, beyond 450\,d
post-transient, we find no confirmed flaring in our sample, and can constrain the luminosity
of such events in nearby LFBOTs to be hundreds of times fainter than the 
original transient. For LFBOTs with $z\gtrsim0.1$, our constraints on
the luminosity of any flaring are roughly comparable to the luminosity of the
original event itself.

We hypothesize that our non-detection of flares in other LFBOTs could
arise from one of two factors: (a) the phenomenon
powering the flares (e.g., accretion) ``shuts off'' or declines $\sim$100\,d after the
initial transient, and (b) the emission from the flares are beamed,
requiring an on-axis observation for a detection. 

We also simulated the appearance of such flares in TESS data, and found
that for a sufficiently nearby LFBOT (occurring once every few years), TESS 
would be able to provide a well-sampled light curve that can be studied
for evidence of late-time flaring. Over the next several years,
the Large Array Survey Telescope (LAST) and ULTRASAT
represent the best facilities with which to 
search for these late-time flares, given their relatively large fields of
view and their high cadences. Interestingly, we also find
that the Rubin/LSST Deep Drilling Fields
could also represent promising avenues for LFBOT flare searches, given
the sub-day cadence and depth.
With LSST, we will also be able to track the evolution of LFBOTs (and their
rates) across cosmic time, and gain a better understanding of their
progenitors and underlying astrophysical processes.

%% Please use the acknowledgment and contribution environments. This will 
%% be anonomyized when the "anonymous" style option is used. 
\begin{acknowledgments}

RJ would like to thank Armin Rest for information about ATLAS
photometry, Eliot Quataert for information about black hole
accretion disk models, and the ZTF publication board for feedback
on this manuscript.

%funding/grants
R.J. would like to acknowledge the support of the Klarman Fellowship. 
A.Y.Q.H. acknowledges support from a Sloan Research Fellowship 
(Award Number FG-2024-21320) from the Alfred P. Sloan Foundation. 
R.K. is grateful for the support of the Dean of Faculty Fellowship.

%tess
This paper includes data collected by the TESS mission.  Funding for the TESS 
mission is provided by the NASA Explorer Program.
The TICA data utilized in this work was 
obtained from the MAST archive at
\dataset[10.17909/t9-9j8c-7d30]{https://dx.doi.org/10.17909/t9-9j8c-7d30},
hosted by the Space Telescope Science Institute (STScI).
STScI is operated by the Association of Universities for 
Research in Astronomy, Inc., under NASA contract NAS 5–26555.

%ztf
This paper is based on data obtained with the Samuel Oschin Telescope 48-inch and the 60-inch Telescope at the Palomar Observatory as part of the Zwicky Transient Facility project. ZTF is supported by the National Science Foundation under Grants No. AST-1440341, AST-2034437, and currently Award \#2407588. ZTF receives additional funding from the ZTF partnership. Current members include Caltech, USA; Caltech/IPAC, USA; University of Maryland, USA; University of California, Berkeley, USA; Cornell University, USA; Drexel University, USA; University of North Carolina at Chapel Hill, USA; Institute of Science and Technology, Austria; National Central University, Taiwan, German Center for Astrophysics, Germany, and OKC, University of Stockholm, Sweden. Operations are conducted by Caltech's Optical Observatory (COO), Caltech/IPAC, and the University of Washington at Seattle, USA.

%ned
This research has made use of the NASA/IPAC Extragalactic Database's extinction
calculator (URL: \doi{10.26132/NED5}), which is funded by the National Aeronautics 
and Space Administration and operated by the California Institute of Technology.

\end{acknowledgments}

%% To help institutions obtain information on the effectiveness of their 
%% telescopes the AAS Journals has created a group of keywords for telescope 
%% facilities.
%
%% Following the acknowledgments section, use the following syntax and the
%% \facility{} or \facilities{} macros to list the keywords of facilities used 
%% in the research for the paper.  Each keyword is check against the master 
%% list during copy editing.  Individual instruments can be provided in 
%% parentheses, after the keyword, but they are not verified.
\facilities{TESS, ZTF, ATLAS, NED}

%% Similar to \facility{}, there is the optional \software command to allow 
%% authors a place to specify which programs were used during the creation of 
%% the manuscript. Authors should list each code and include either a
%% citation or url to the code inside ()s when available.
\software{{\tt astropy} \citep{astropy:2013,astropy:2018,astropy:2022}, 
{\tt astroquery} \citep{astroquery},
{\tt numpy} \citep{numpy}, {\tt matplotlib} \citep{matplotlib}, {\tt tess-point} \citep{tess_point}, {\tt lightkurve} \citep{lightkurve}, {\tt scipy} \citep{scipy},
{\tt synphot} \citep{synphot}}

%% Appendix material should be preceded with a single \appendix command.
%% There should be a \section command for each appendix. Mark appendix
%% subsections with the same markup you use in the main body of the paper.
%%
%% Each Appendix (indicated with \section) will be lettered A, B, C, etc.
%% The equation counter will reset when it encounters the \appendix
%% command and will number appendix equations (A1), (A2), etc. The
%% Figure and Table counter will not reset.
\appendix

\section{Statistically ruling out late-time detections as flares}
\label{app:stat_tests}

\begin{figure*}
    \centering
    \includegraphics[width=\linewidth]{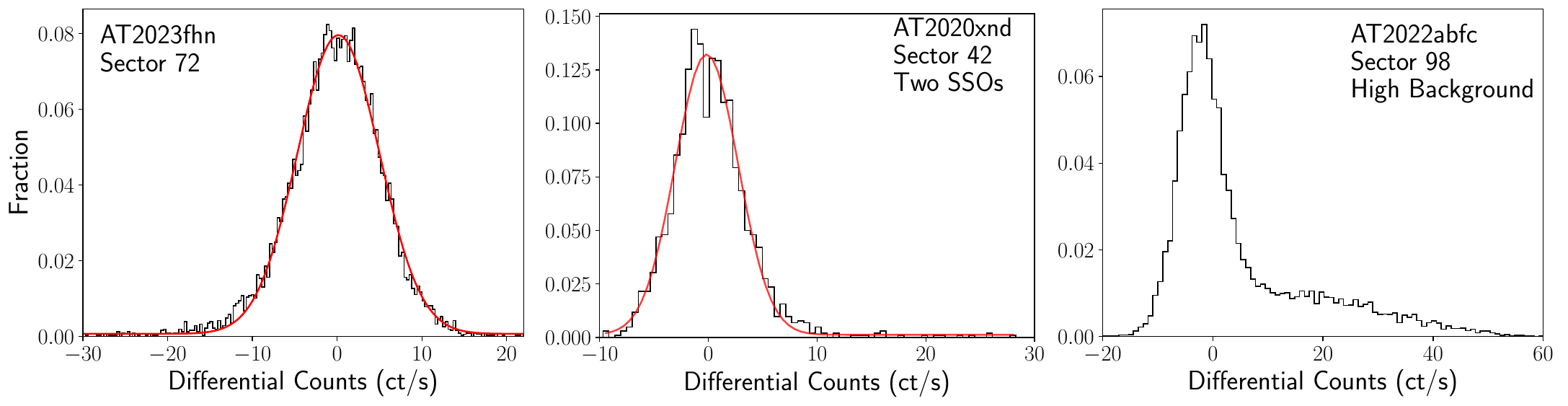}
    \caption{Flux distributions for three sample light curves from the ones we analyzed in our
    sample, with fits to Gaussians in two cases. \textit{Left}: AT2023fhn shows a featureless
    light curve whose flux distribution can be well approximated by a Gaussian. \textit{Center}: AT2020xnd
    had two SSOs pass through the aperture in Sector 42, which leads to outliers in the distribution
    from the excess flux; the rest of the distribution is well-approximated by a Gaussian. \textit{Right}:
    AT2022abfc's Sector 98 light curve suffered from significant scattered light, leading to a skewed
    distribution that is not approximable using a Gaussian.}
    \label{fig:flux_dists}
\end{figure*}

Late-time ZTF detections were ruled out for the first six LFBOTs in our sample by 
\citetalias{ho_tsd_flares}. In this Appendix, we rule out any detections as arising from
flaring using the TESS data from all LFBOTs, as well as the ZTF data from the most recent six
FBOTs in our sample. For a detection to be statistically significant, we impose the
condition that it must have a chance occurrence probability of $<1\%$. None of the
late-time detections that we find here have such a low probability of chance occurrence.
Figure \ref{fig:flux_dists} shows example flux distributions from the
TESS difference images for three of the most common
kinds of light curves (featureless, with SSOs, or high levels of scattered light); in
cases where scattered light was not present, we can use Gaussian
statistics to estimate probabilities of spurious detections.

We also direct the reader to \citet{ofek_wpp_limits} for the details of a targeted 
search using the Large Array Survey Telescope (LAST) to identify AT2022tsd-like flaring in AT2024wpp.
Details about targeted, high-cadence follow-up of AT2026dbl will be presented in Z. McGrath et al., in prep.

\paragraph{CSS161010} In TESS, there was one spurious detection in Sector 5, which has a probability
of chance occurrence of roughly 79\%. In Sector 32, there is 1 spurious detection; the probability
of such a detection is over 99\%. There are no spurious late-time detections in Sector 98.

\paragraph{AT2018cow} In TESS, Sector 25 suffered from severe scattered light toward the end of 
each of the two orbits. As a result, there are several spurious detections where the background
was difficult to model, and these detections during periods of high scattered light should be 
discounted. For Sectors 51--52, which also suffered from significant scattered light, a large
fraction of the data was unusable. For the portion of the data that we did use, we saw 3 detections,
which had a chance occurrence probability of roughly 85\%. These were one-off points with elevated
flux that occurred during times of elevated background.
For Sector 78, there is 1 detection; this has a probability of chance occurrence of over
99\% percent. As a result, none of these detections are likely real.

\paragraph{AT2018lug} In TESS Sectors 42--43, there are no spurious detections; the SSO moving through the
aperture was discussed in Section \ref{sec:results}. In Sectors 70--71, there are 8 formal detections
in over 17000 photometric flux measurements; this has a chance of occurring of over 99\%.

\paragraph{AT2020mrf} In TESS Sectors 50--51, there
are 3 points with detections above the magnitude limit, though these correspond to epochs with a 
high level of background due to scattered light. The probability of chance occurrence of these detections is 98.5\%, 
suggesting that they are not real. In Sectors 77--78, there are several spurious detections due 
to elevated background; there is no coherent signal in this light curve. 
In particular, the background begins to exhibit a periodic variation toward
the end of the observation.

\paragraph{AT2020xnd} The Sector 42 measurements suffer from significant scattered light, as well 
as two SSOs passing through the aperture (see Section \ref{sec:results}). There are only 2 spurious
detections outside of these times, which would occur due to pure chance at an 82\% probability. In 
Sector 55, there is a significant time-varying background, which could explain most of the spurious
detections. In Sector 92, there are 2 detections, which could occur due to random chance with
a probability of $>99\%$.

\paragraph{AT2022tsd} The TESS observations contain two points that are formal detections, apart from
the SSO moving through the aperture (described in Section \ref{subsec:ssos}). There is a very high 
likelihood (nearly 100\%), out of over 19\,000 observations, that there could be 2 spurious 
detections. Additionally, these occur during regions of high background, which suggests they are spurious.
From ZTF, there is one detection (S/N 3.3) roughly 800 days after the last detection (at roughly MJD 60690), 
a few years after the last flares were observed. The probability of this happening 
due to pure chance is roughly 13\%, suggesting that this is not real. Inspecting the ZTF image
does not reveal evidence for a source there.

\paragraph{AT2022abfc} There are no late-time detections in ZTF, though there is a late-time detection in
ATLAS photometry at MJD 60283.18512 (roughly $t_0+380$\,d), with a 
significance of 4.6-$\sigma$. While this detection is not 
obviously an artifact, it does not have a clear PSF-like shape; the ATLAS PSF is undersampled, so this
detection being real is certainly a possibility. If real, the measured flux is $30\pm7$\,$\mu$Jy; at this
transient's redshift of 0.212, this corresponds to a luminosity of $(1.9\pm0.5)\times10^{43}$\,erg\,s$^{-1}$. 
The original transient had a peak AB magnitude of $19.1\pm0.1$ in ATLAS, corresponding to a peak 
luminosity of $5.7\times10^{43}$\,erg\,s$^{-1}$. If this were a \textit{bona fide} flare from 
AT2022abfc, the luminosity would comparable to those observed from AT2022tsd \citepalias{ho_tsd_flares}.
The last $r$-band ZTF upper limit is 19.05, 15 days prior, and the subsequent upper limit is
19.29, roughly five days later, making it difficult to determine if this detection is real. 
Moreover, publicly-available Pan-STARRS1 photometry \citep{ps1} does not extend till this epoch.

In TESS observations from Sector 98, there are several spurious
detections that are strongly correlated with regions of the light curve with elevated background,
suggesting that these are not real.

\paragraph{AT2023fhn} We found a single detection in ZTF with an SNR of 3.1. Out of 453 images, the chance of
exactly 1 image yielding such a detection is $\sim28\%$. Inspection of the ZTF image at this epoch did
not show evidence for a clear point source. There were no significant detections in TESS throughout
the duration of the Sector 72 observations.

\paragraph{AT2023hkw} We found a single detection in ZTF with an SNR of 3.2. Out of 650 images, the chance
of exactly 1 image yielding such a detection is roughly 28\%, suggesting that this detection is 
spurious. Inspection of this ZTF image revealed significant contamination from satellite tracks (see, e.g.,
\citealt{mroz_starlink}), one of which
passed through the location of the transient at this image. In TESS, there is one formal 
detection out of 7\,681 photometric points, which has a chance probability of occurring of over 99\%.

\paragraph{AT2023vth} We did not find any late-time detections from ZTF, although there was a 
late-time detection in ATLAS. We inspected the ATLAS difference image at this time using the online photometry service
and found an artifact in the data, with a black strip running through the center of the 
image (through the location of the LFBOT)---making this likely a spurious detection. 
In TESS, there are several detections that occur during periods of high
background; the temporal profile of these detections is correlated with the changes in background,
suggesting these are systematic artifacts rather than real detections.

\paragraph{AT2024qfm} We found one late-time detection in ZTF with an SNR of 3.7 out of 217 images. 
There is a $\sim2\%$ chance of this occurring purely via chance. We inspected the difference image at
this position and found no evidence for a source. In addition, there is also an ATLAS detection at MJD 
60540.43737, although inspection of the difference images reveals that this could arise from spurious
photometry due to a black band running through the bottom of the image, almost through the location
of the LFBOT. In TESS, there is one detection out of
over 10\,000 photometric points, which has a probability of chance occurrence of nearly 100\%, suggesting that
this detection is not real. 

\paragraph{AT2024wpp} There are 3 detections in TESS Sector 97 out of a total of 
9\,416 points, which has a probability of chance occurrence of almost 100\%. This suggests that 
these detections are not real. There are no late-time detections after the last detection in ZTF.

\paragraph{AT2024aehp} There is no post-transient TESS data. There are also no detections from ZTF
after the last detection by ZTF that is associated with the original transient.

\paragraph{AT2026dbl} We found one late-time detection in ZTF with an SNR of 3.3, out of 95 images.
There is a $\sim$4.7\% chance of this occurring purely via chance, suggesting that this detection
is spurious. There is no evidence for a source in the corresponding 
ZTF difference image at this position.

\bibliography{lfbot_limits}{}
\bibliographystyle{aasjournalv7}

%% This command is needed to show the entire author+affiliation list when
%% the collaboration and author truncation commands are used.  It has to
%% go at the end of the manuscript.
%\allauthors

%% Include this line if you are using the \added, \replaced, \deleted
%% commands to see a summary list of all changes at the end of the article.
%\listofchanges

\end{document}